\documentclass[11pt]{article}
\usepackage{graphicx}
\usepackage{float} %
\usepackage{ifthen}

\usepackage{pifont}
\usepackage{latexsym}
\usepackage{amsfonts}
\usepackage{amsmath}
\usepackage{amssymb}

\makeatletter
\def\@citex[#1]#2{\if@filesw\immediate\write\@auxout{\string\citation{#2}}\fi
  \def\@citea{}\@cite{\@for\@citeb:=#2\do
    {\@citea\def\@citea{,\linebreak[0]}\@ifundefined {b@\@citeb}{{\bf
          ?}\@warning {Citation `\@citeb' on page \thepage \space
          undefined}}%
      \hbox{\csname b@\@citeb\endcsname}}}{#1}}

\let\shortcite\cite
\newcommand{\cucm}[1]{\ensuremath{\mbox{\rm{}#1}\hbox{-}\allowbreak\mbox{\rm{}CUCM}}}
\newcommand{\cwcm}[1]{\ensuremath{\mbox{\rm{}#1}\hbox{-}\allowbreak\mbox{\rm{}CWCM}}}
\newcommand{\cucmrunoff}[1]{\ensuremath{\mbox{\rm{}#1}\hbox{-}\allowbreak\mbox{\rm{}CUCM}\hbox{-}\allowbreak{}\mbox{\rm{}runoff}}}
\newcommand{\cwcmrunoff}[1]{\ensuremath{\mbox{\rm{}#1}\hbox{-}\allowbreak\mbox{\rm{}CWCM}\hbox{-}\allowbreak{}\mbox{\rm{}runoff}}}
\newcommand{\cucmrevoting}[1]{\ensuremath{\mbox{\rm{}#1}\hbox{-}\allowbreak\mbox{\rm{}CUCM}\hbox{-}\allowbreak{}\mbox{\rm{}revoting}}}
\newcommand{\cwcmrevoting}[1]{\ensuremath{\mbox{\rm{}#1}\hbox{-}\allowbreak\mbox{\rm{}CWCM}\hbox{-}\allowbreak{}\mbox{\rm{}revoting}}}

\newcommand{\twthen}{\mbox{\sc Then}}
\newcommand{\xthenx}[1]{{#1} {\twthen} {#1}}
\newcommand{\xtheny}[2]{{#1} {\twthen} {#2}}
\newcommand{\xthenxrevoting}[1]{\xthenx{#1}\ (with revoting)}
\newcommand{\p}{{\rm P}}
\newcommand{\np}{{\rm NP}}
\newcommand{\elec}{\electionsystem}
\newcommand{\electwo}{\electionsystemtwo}
 \newcommand{\elece}{{\cale}}
\newcommand{\electionsystem}{\italx}
\newcommand{\electionsystemtwo}{\italy}
\newcommand{\cale}{\ensuremath{\cal E}}
\newcommand{\italx}{\ensuremath{X}}
\newcommand{\italy}{\ensuremath{Y}}

\newlength{\filength}
\newsavebox{\gcbox}
\sbox{\gcbox}{\framebox[\filength]{\rule{0ex}{2ex}}}

\newtheorem{theorem}{Theorem}[section]

\newcommand\qedblob{\ding{113}}
\def\literalqed{{\ \nolinebreak\hfill\mbox{\qedblob\quad}}}

\newtheorem{definition}[theorem]{Definition}
\showhyphens{}
\newcommand{\singlespacing}{\let\CS=
\@currsize\renewcommand{\baselinestretch}{1}\tiny\CS}
\newcommand{\singlespacingepsilon}{\let\CS=
\@currsize\renewcommand{\baselinestretch}{1.05}\tiny\CS}
\newcommand{\singlespacingplus}{\let\CS=
\@currsize\renewcommand{\baselinestretch}{1.25}\tiny\CS}
\newcommand{\doublespacing}{\let\CS=
\@currsize\renewcommand{\baselinestretch}{1.75}\tiny\CS}
\newcommand{\extradoublespacing}{\let\CS=
\@currsize\renewcommand{\baselinestretch}{1.9}\tiny\CS}
\newcommand{\nicenicespacing}{\let\CS=
\@currsize\renewcommand{\baselinestretch}{1.9}\tiny\CS}
\newcommand{\draftspacing}{\let\CS=
\@currsize\renewcommand{\baselinestretch}{2.0}\tiny\CS}
\newcommand{\hugedraftspacing}{\let\CS=
\@currsize\renewcommand{\baselinestretch}{2.4}\tiny\CS}
\newcommand{\niceonespacing}{\let\CS=\@currsize\renewcommand{\baselinestretch}{1.1}\tiny\CS}
\newcommand{\nicetwospacing}{\let\CS=\@currsize\renewcommand{\baselinestretch}{1.2}\tiny\CS}
\newcommand{\nicethreespacing}{\let\CS=\@currsize\renewcommand{\baselinestretch}{1.3}\tiny\CS}
\newcommand{\singlespacingplusplus}{\let\CS=\@currsize\renewcommand{\baselinestretch}{1.35}\tiny\CS}
\newcommand{\nicefourspacing}{\let\CS=\@currsize\renewcommand{\baselinestretch}{1.4}\tiny\CS}
\newcommand{\nicefivespacing}{\let\CS=\@currsize\renewcommand{\baselinestretch}{1.5}\tiny\CS}
\newcommand{\nicesixspacing}{\let\CS=\@currsize\renewcommand{\baselinestretch}{1.6}\tiny\CS}
\newcommand{\nicesevenspacing}{\let\CS=\@currsize\renewcommand{\baselinestretch}{1.7}\tiny\CS}
\newcommand{\niceeightspacing}{\let\CS=\@currsize\renewcommand{\baselinestretch}{1.8}\tiny\CS}
\newcommand{\niceninespacing}{\let\CS=\@currsize\renewcommand{\baselinestretch}{1.9}\tiny\CS}
\makeatother%



\newcommand{\sat}{{\rm SAT}}

\newcommand{\npc}{\np\text{-complete}}
\newcommand{\npccc}{\ensuremath{{\rm NPC}}}

\newenvironment{proofs}{\noindent{\bf Proof.}\hspace*{1em}}{\literalqed\bigskip}
\newenvironment{proofsketch}{\noindent{\bf Proof Sketch.}\hspace*{1em}}{\literalqed\bigskip}

\DeclareMathSymbol{\subsetneq}{\mathbin}{AMSb}{"28}
\DeclareMathSymbol{\supsetneq}{\mathbin}{AMSb}{"29}

\newcommand{\pair}[1]{\mathopen\langle{#1}\mathclose\rangle}
\newcommand{\condition}{\,\mid \:}
 
\def\land{{\; \wedge \;}}
\def\lor{{\; \vee \;}}

\newcommand{\numvars}{{\rm \#vars}}

\newcommand{\alwayswinners}{\mathit{AlwaysWinners}}

\title{Manipulation Complexity of Same-System Runoff Elections\thanks{Supported in part by NSF grants
    CCF-0915792, CCF-1101452, and CCF-1101479, and
NSF Graduate Research Fellowship DGE-1102937. 
}} 

\author{Zack Fitzsimmons\\
College of Computing and Inf.\ Sciences\\
  Rochester Inst.\ of Technology \\
  Rochester, NY 14623 \and
  Edith Hemaspaandra \\
  Dept.~of Computer Science\\
  Rochester Inst.\ of Technology \\
  Rochester, NY 14623 \and
  Lane A. Hemaspaandra\\
  Dept.~of Computer Science\\
  University of Rochester \\
  Rochester, NY 14627 }

\date{June 20, 2014}

\begin{document}
\sloppy 

\maketitle

\begin{abstract}
  Do runoff elections, using the same voting rule as the initial
  election but just on the winning candidates, increase or decrease
  the complexity of manipulation?  Does allowing revoting in the
  runoff increase or decrease the complexity relative to just having a
  runoff without revoting?  
  For both weighted and
  unweighted voting, we show that even 
  for election systems with simple winner problems the complexity of
  manipulation, manipulation with runoffs, and manipulation with
  revoting runoffs are 
  independent, in the abstract.
  On the other hand, for 
some important, 
 well-known election systems we
  determine what holds for each of these cases.
For no such systems do we find runoffs lowering complexity, and for 
some we find that runoffs raise complexity.
Ours is the first paper to show that for natural, unweighted election
systems, runoffs can increase the manipulation complexity.
\end{abstract}

\section{Introduction}
There is an extensive literature on two-stage and multistage voting.
Although some of this study exists within economics, multistage elections and
runoffs have been greatly influential in computational social choice
during the past decade, due to such work as that 
of Elkind and Lipmaa~\shortcite{elk-lip:c:hybrid-manipulation}
and 
Conitzer and Sandholm~\shortcite{con-san:c:nonexistence}.

Particularly interesting recent
work in this line 
has been done
by 
Narodytska and Walsh~\shortcite{nar-wal:c:two-stage}.
They focus on manipulation of election
systems of the form \xtheny{$X$}{$Y$}, i.e., an initial-round election
under voting rule $X$, after which if there are multiple winners just
those winners go on to a runoff election under voting rule $Y$,
with the initial votes now restricted to the remaining candidates.
  The
question at issue is whether a given manipulative coalition can vote
in such a way as to make a distinguished candidate win (namely, win in
the initial round if there is a unique winner in the initial round, or if
not, then be a winner of the runoff).  

Narodytska and Walsh~\shortcite{nar-wal:c:two-stage}
study the
computational complexity of this question.
They strongly address the issue of how the manipulation complexity of
$X$ and $Y$ affect the manipulation complexity of \xtheny{$X$}{$Y$}.
Viewing P as being easy and NP-hardness as being hard,
they show that every possible combination of these manipulation
complexities can be achieved for $X$, $Y$, and \xtheny{$X$}{$Y$}.

The present paper focuses on the complexity of \xthenx{$X$}.  That is,
we are focused on the case where $X$ is so valued as an election
system that if $X$ selects a unique winner, our election is over and
we have our winner.  However, if $X$ in the initial round 
has tied winners,
then we take just those winners and subject them to
a runoff election, again using system $X$.  (Votes in this second
election will be over only the candidates who made it to the second
round.)
Like Narodytska and Walsh~\shortcite{nar-wal:c:two-stage},
we are interested in the case in which the second-round 
votes are simply the initial-round votes restricted to the remaining 
candidates, and the case, first raised 
by them,
in which revoting is allowed in the second
round.

Real-world examples exist of such same-system runoff elections.
In 
general
elections in North Carolina and many
districts of California, election law specifies that if there are two
or more candidates tied for being the winner
in the initial plurality election, a 
plurality runoff election is held among just those 
candidates~\cite{north-carolina:law:ties,california:law:ties-in-elections}.
(Curiously, in both states this approach is explicitly limited to
general elections.  For party-candidate-selection (so-called ``primary'') elections, perhaps to 
limit cost,
both states break ties by lot, and in addition North Carolina breaks
ties by lot if very few
voters voted.)
So (\xthenx{Plurality})-with-revoting is being used.

Although 
Narodytska and Walsh~\shortcite{nar-wal:c:two-stage}
for \xtheny{$X$}{$Y$} elections
showed that all combinations of P and NP-hardness 
for $X$, $Y$, and \xtheny{$X$}{$Y$} 
can be realized,
their examples achieving that almost all have $X \neq Y$. Thus
their broad results do not address the issue of whether all
possibilities can be achieved if one seeks to use the same system
for both the initial and the runoff election.  
We show that every possibility can be achieved, even when
the runoff is the same system as the initial election.  Indeed, even
in the three-way comparison of the complexity of $X$, the complexity
of $X$ with runoff (under $X$), and the complexity of $X$ with a runoff 
(under $X$) with
revoting, we show that every possibility of setting some or all of
those to P or to NP-complete manipulation complexities can be
realized.  And we show that this can even be done while ensuring that the
winner problem for $X$ (i.e., determining whether a given candidate is 
a winner of a given election under $X$) 
remains in P, and can also be done both for the
weighted and the unweighted cases.  For example, there are election
systems $X$---having $\p$ winner problems---such that manipulation
of $X$ is NP-complete, manipulation of \xthenx{$X$} is NP-complete,
but manipulation of \xthenx{$X$} with revoting is in $\p$.  And there
are election systems $X$---having $\p$ winner problems---such
that manipulation of $X$ is in P, manipulation of
\xthenx{$X$} is NP-complete, but manipulation of \xthenx{$X$} with
revoting is in $\p$.  Briefly put, there is no inherent 
connection between these three complexities.

For the most important systems, however, it is very important to see
what the effects of runoffs, and revoting runoffs are.  For example,
weighted plurality is easily seen to remain easy in all of our cases,
e.g., manipulation of elections with runoffs, or with revoting
runoffs, remains in $\p$.  However, that result itself is something of
a fluke.  We show that for every (so-called) scoring protocol that is
not Triviality, Plurality, or a disguised version of one of those,
manipulation of elections with runoffs and manipulation of elections
with revoting runoffs are NP-complete.  Although manipulation of
unweighted veto is in P, we show that manipulation of unweighted 
veto elections
with runoffs and manipulation of unweighted 
veto elections with revoting runoffs
are NP-complete.  For unweighted HalfApproval (the scoring
protocol where each voter gives one point to his or her 
$\lceil \|C\|/2
\rceil$ top candidates and zero points to the rest), we prove
that for both elections with runoffs and elections with revoting
runoffs, the manipulation complexity,
even when
restricted to having at most one manipulator,
is NP-complete.
This contrasts with the
nonrunoff manipulation complexity 
of unweighted HalfApproval, which is in P
when there is one manipulator (and indeed for an unlimited 
number of manipulators, using the argument
of~\cite[Corollary 4.2]{con-pro-xia:c:scheduling-manipulation}).

Both Veto and HalfApproval are natural, unweighted cases where runoffs
increase the manipulation complexity, and are the first such natural,
unweighted cases in the literature.  In fact, the previous literature 
did not find any such natural, unweighted example 
\emph{even if one is allowed to use different election systems 
in the first and second rounds}.
And for weighted manipulation, we provide the literature's first
natural examples where a same-system runoff increases manipulation
complexity.

For the case of one manipulator, 
a standard way of
seeking to
manipulate unweighted or weighted scoring protocols---pioneered for
the unweighted case by 
Bartholdi, Tovey, and Trick~\shortcite{bar-tov-tri:j:manipulating}, 
and
extended in many papers since---is to use the natural greedy
algorithm.  However, we prove that for some scoring protocols $X$, the
greedy approach fails on \xthenx{$X$}.

Since we feel this tool will be useful elsewhere, we mention that
Section~\ref{sec:alwayswinners} presents a tool, which we call the
$\alwayswinners$ transformation, that is very helpful in paving over
the chasm caused by the fact that some people feel that the natural
way to define the notion of an election rule is to allow any subset of the
candidates to form the winner set, and some people feel that the
natural way to define the notion of an election rule is to allow any
\emph{nonempty} subset of the candidates to form the winner set.
Traditional social choice uses the latter definition, but many
computationally oriented papers prefer the former for its symmetry.
Section~\ref{sec:alwayswinners}'s $\alwayswinners$ 
tool is a construction that 
transforms an election system into a new one that will (except when the
candidate set is empty) always have at least one winner.  Crucially,
the transformation is so tightly related to the original election
system that it leaves unchanged the complexity of many
election-attack problems.
\section{Related Work}
There are quite a few papers whose focus is close to ours.  Yet each
differs in some important 
way.  
Centrally
underpinning our study and framing is the 
work of 
Narodytska and Walsh~\shortcite{nar-wal:c:two-stage}
on
manipulating \xtheny{$X$}{$Y$} elections.
In a very real sense, our
paper is merely about their diagonal---the case when one uses the same
voting system in the original election and the runoff.  However, since they
were not specifically exploring the diagonals, their existence results
in general don't address that case.
(However, we must mention an
important exception.  They show that for STV$'$, a particular decisive 
form of STV, that STV$'$ and \xthenx{STV$'$} are both 
NP-hard.%
\footnote{To avoid confusing the literature's terminology, 
it is important for us to mention that there is a very slight,
but arguably philosophically interesting, difference
between the $\twthen$ we defined in the Introduction and the $\twthen$
operator as defined by 
Narodytska and Walsh~\shortcite{nar-wal:c:two-stage}.
Our and their definitions 
of \xtheny{$X$}{$Y$}
can differ in outcome only
on what happens if there is exactly one winner of the initial 
election.  In our use of \twthen\ (as given in this paper), 
in that decisive case the election is 
over.  In their case, that one winner goes on to a one-person 
election under system $Y$.  Their approach opens the door to 
having system $Y$ in some cases kill off a single candidate 
who won the initial round.  However, we stress that in their 
paper they absolutely never use that possibility, and so 
every result in their paper, including each one mentioned 
in this paper, holds equally well in both models.  Indeed, for 
any election system that always has at least one winner when
there is at least one candidate, the two models coincide, and 
almost all natural election systems have this property.
}%
)
Our constructions, which must work within a 
single system for both rounds,
are quite different from theirs.

In contrast, the even earlier work of 
Elkind and Lipmaa~\shortcite{elk-lip:c:hybrid-manipulation} 
has a
section on using the same system in each round, which is our focus
also.  However, their model (unlike 
Narodytska and Walsh
and unlike our paper, which pass forward just the winners) is based on
removing only the \emph{least} successful 
candidate after a round.  In
particular, their model is of one or more initial pruning rounds, that 
in their examples use a
``prune off the least successful candidate'' (except in one case they
prune off half the candidates) rule inspired by some election system $X$,
after which there is a final round using some (potentially
different) election system $Y$.  So
Elkind and Lipmaa's 
section on using the same system even in the final
round (Section~5 of their paper) 
is about having one or more
rounds using (a variant of) $X$ to cut off the least popular
candidate, and then a final round also using $X$.  Other recent 
work on removing weakest candidates, usually sequentially, 
include that of 
Bag, Sabourian, and Winter~\shortcite{bag-sab-win:j:sequential-elimination}
and
Davies, Narodytska, and Walsh~\shortcite{dav-nar-wal:c:weakest-link}.

Related to the Elkind--Lipmaa 
work is the 
still earlier 
``universal tweaks'' work of
Conitzer and Sandholm~\shortcite{con-san:c:voting-tweaks}, 
which shows that adding one
pairwise (so-called) CUP-like ``preround,'' which cuts out about half
the candidates, can tremendously boost a system's manipulation
complexity over a broad range of systems.

Speaking more broadly, the problem that 
Narodytska and Walsh~\shortcite{nar-wal:c:two-stage}
and this paper are studying, for the case of runoffs and runoffs with
revoting, is the manipulation problem.  This asks whether a coalition
of manipulators can ensure that a particular candidate is a winner of
the overall election.  The seminal work on the computational
complexity of manipulation was that
of 
Bartholdi, Tovey, and Trick~\shortcite{bar-tov-tri:j:manipulating}
and 
Bartholdi and Orlin~\shortcite{bar-oli:j:polsci:strategic-voting}, 
and there have been
many papers since studying manipulation algorithms for, and hardness
results for, a variety of election systems, see, e.g., the
survey~\cite{fal-hem-hem-rot:b-too-short:richer}.  This entire stream
exists within the area known as computational social
choice~\cite{che-end-lan-mau:c:polsci-intro}.

Finally, we mention that there is an interesting line of work of Meir 
et al.~\shortcite{jen-mei-pol-ros:c:iterative-plurality}
and 
Lev and Rosenschein~\shortcite{lev-ros:c:iterative-voting}
studying in a fully game-theoretic setting 
iterated 
voting
in the sense of seeing whether a Nash equilibrium is 
reached; 
see also Reijngoud and Endriss~\shortcite{end-rei:c:iterated-polls}.
This work does not remove candidates after votes, and
so is different in flavor and goal from our work.

\section{Preliminaries}
We first give a standard formalization of elections, the winner
problem, and the manipulation problem.  
Each election instance will
have a finite set, $C$, of candidates, e.g., a particular election
might have Obama and Romney as its candidates.  Elections also have a
finite collection 
of votes, which we will assume are input as a list of ballots, one
per voter.  
Although social choice theory sometimes allows voters to
have names, in this paper we study the most natural case---the one
where votes come in nameless, and the election system's outcome
depends on just what the multiset of votes is.
We will 
refer to 
the collection of votes as $V$. 
The type of each vote will depend on the election system.  Most
systems require 
a tie-free linear ordering of the
candidates, and that will be the case for all systems discussed in
this paper. 

So-called scoring protocols such as Plurality, Veto, Borda, and so on will for
us have votes cast as linear orders.  And then from those orders we
will assign points to each candidate based on the rules of that
scoring system.  For example, 
in a veto election, each voter casts
zero points for his or her least favorite candidate, and one point for
each other candidate.  
In a plurality election, each voter casts
one point for his or her favorite candidate, and zero points for
each other candidate.  
In Triviality, each candidate gets zero %
points from each
voter, and so all candidates tie as winners.
In HalfApproval, if there are $m$ candidates,
each voter gives one point to each of the $\lceil m/2 \rceil$ top
candidates in his or her linear order, and gives zero points to each
other candidate.  
In any scoring system, all points for each candidate
are added up, and the candidate(s) who have the maximum score achieved
by any candidate are the winner(s).
(When we speak of scoring protocols in the abstract, 
each scoring protocol must have a fixed number of candidates.
However, when we say Plurality or HalfApproval or so on, we usually are 
referring to the protocol that on $m$-candidate inputs uses 
the $m$-candidate Plurality or HalfApproval or so on scoring 
protocol mentioned above.)
In Copeland$^\alpha$ 
elections~\cite{cop:unpub:copeland}, where $\alpha$ is a rational number
between $0$ and $1$, winners are computed as follows.  We look at the
pairwise election between every pair of candidates.  The candidate that
wins gets a point.  In case of a tie, both candidates get $\alpha$
points. The winner(s) are the candidates with the most points.
Copeland$^1$ is also known as 
Llull~\cite{hae-puk:j:electoral-writings-ramon-llull}.  

An election system, $\electionsystem$, is a mapping that given
$C$ and $V$ outputs 
a set of candidates $W$ (``the winner(s)'').
This 
is precisely 
the definition of a social choice
correspondence, as given in
Shoham and 
Leyton-Brown~\shortcite{ley-sho:b:multiagent-systems}.\footnote{That 
  definition and
  this paper allow, as do many papers in computational social choice
  theory, the case in which an election has no winners.  We find that 
  natural for symmetry with the case in which everyone wins.  Also, 
  there are real-world cases in which having no winner is natural, e.g., the 
  system for electing players to the Baseball Hall of Fame is set up
  so that if the crop of candidates in a given year is weak no 
  one will win.  That has happened four times, most
  recently in the January 2013 vote, in which none of the 37 candidates 
  were elected to the Hall.}
All 
election systems that we construct to realize claims of our theorems have 
$W\neq \emptyset$ (when $C \neq \emptyset$),
which in social choice is often part of the definition of an
election system (see also the paragraph immediately before the start
of Section~\ref{sss:rt}).
Our Section~\ref{sec:alwayswinners}
in fact gives a broadly
applicable way of paving over the difference between having and not having 
this condition.
The winner problem for an election system $\elec$ is the language 
that contains exactly those triples $C$, $V$, and $p\in C$ such that 
$p$ is a winner in the $\elec$ election on $C$ and $V$.  
Although some well-known election systems exist whose winner 
problems are not in P~\cite{bar-tov-tri:j:who-won}, 
all the systems we study in this paper 
have $\p$ winner problems.

We now define the classic unweighted and weighted (coalitional) election 
manipulation problems, due to 
Conitzer, Sandholm, and Lang~\shortcite{con-lan-san:j:when-hard-to-manipulate}
(generalizing 
the noncoalitional 
unweighted case 
raised by Bartholdi, Tovey, and Trick~\shortcite{bar-tov-tri:j:manipulating}).
The unweighted version, called Constructive Unweighted 
Coalitional Manipulation (CUCM), is defined as follows for 
any given election system $\electionsystem$.

\begin{description}

\item[Name:] $\cucm{\electionsystem}$.

\item[Given:] A set $C$ of candidates, a 
collection $V_1$ of the nonmanipulative votes (each specified by 
a tie-free linear ordering over the candidates),
   a set $V_2$ of manipulative 
voters %
(since our voters do not have names,
these
are specified by a nonnegative integer input in unary 
giving the number of manipulative voters), and 
    a distinguished candidate $p \in C$. 

 \item[Question:] Is there a way to
set the votes of the manipulators, $V_2$, so that 
   under the election system~$\elec$, $p$ is a winner of
   the election over candidate set $C$ with the vote set being 
   the ballots of the manipulators and the nonmanipulators?
\end{description}

The analogous weighted version,
\cwcm{\elec},
 is the same except each member of $V_1$
has both a weight and a tie-free linear order, and $V_2$ is specified as
a list giving the weight of each manipulator.  The allowed range of 
weights is the positive integers.
For each of the election systems we deal with in a weighted context,
it will be immediately apparent what it means to use the election system 
on weighted voters.

Our interest here is in runoff elections.  So in addition to the
above classic versions, let us define versions with runoffs and with
revoting runoffs.  The ``runoff'' problems \cucmrunoff{\elec} and
\cwcmrunoff{\elec}\ are the same as the above problems, except if
after the $\elec$ election there are two or more winners, a runoff
election is conducted under $\elec$, with the candidates being just the
winners of the initial election, and the votes of all voters (both
manipulators and nonmanipulators) being their initial-election's
preference-order vote, restricted to the remaining set of candidates.
The ``revoting runoff'' (or for short, ``revoting'') problems
\cucmrevoting{\elec} and \cwcmrevoting{\elec}\ are the same as the
above runoff problems, except if there is a runoff election, the
manipulators may change their votes.  And the question is, of course,
whether in this setting there is a set of initial-round and, if needed,
second-round manipulator votes that makes $p$ a winner of the overall
election.\footnote{%
To make this crystal clear,
we give an in-math 
definition
of the $X$-CUCM-runoff notion described in words above, namely, 
given 
$C$, $V_1$, $V_2$, and $p$, the question is whether
$(\exists R) [ \{p\}= X(C, V_1 \cup R) \lor (\| X(C, V_1 \cup R)\| > 1
\land p \in X(X(C, V_1 \cup R), (V_1 \cup R)_{X(C, V_1 \cup R)}))]$,
where the subscript denotes masking those votes down to the specified
candidates, the unions are multiset-like, and $R$ is an
assignment of votes to the manipulative voter set $V_2$.
And similarly, 
$X$-CUCM-revoting is the same except with the question (and the 
new variable $S$ is over assignments of manipulative voters to
$V_2$) being 
$(\exists R)(\exists S) [ \{p\}= X(C, V_1 \cup R) \lor (\| X(C, V_1 \cup R)\| > 1
\land p \in X(X(C, V_1 \cup R), (V_1 \cup S)_{X(C, V_1 \cup R)}))]$.
}
Note that all of these problems are defined as language problems, as
is standard in the area.  
Typical complexities that they might take on are
membership in P and NP-completeness.  Those two cases are
the focus of this paper and of most papers in this area.  However, we
mention in passing three related issues.  First, it has recently been pointed
out that at least in some artificial cases, election decision problems
can be in P even when their related search problems are
NP-hard~\cite{hem-hem-men:c:search-versus-decision}.  This
worry does not infect any of this paper's results.  Every result where
we make a polynomial-time claim in this paper has the property that in
polynomial time one can even produce the action(s) that achieve the
desired outcome (such as making the given candidate win), i.e., our
polynomial-time results are essentially what is sometimes called
``certifiable,'' see
Hemaspaandra, Hemaspaandra, and Rothe~\shortcite{hem-hem-rot:j:destructive-control}.\footnote{For 
the case of revoting runoffs, the natural
  model here, in terms of seeking a polynomial-time certificate-
(i.e., action-) yielding
  algorithms,
 is to
  allow the manipulative coalition, before the runoff election, a full
  view of all the initial votes and candidates, and of the outcome of
  that election, and to require that they set their votes in polynomial time,
  and of course to also require that their initial-election vote-setting
  be done in polynomial time.  However, 
  since all the election systems in this paper have
  polynomial-time winner problems, after a given set of initial-round
  votes the manipulators can themselves compute who the initial-round 
  winner(s) are, and so for problems with p-time winner algorithms,
  one can w.l.o.g.\ require the manipulative coalition to fork over
  at the same time both of its rounds of votes.}
Second, and on
the other end of the complexity range, there
has been much worry about,
and some empirical studies 
suggesting, that 
perhaps even NP-complete sets can be often easy.  
Only during the past half decade has 
computer science obtained the following remarkably strong
result showing that this cannot happen:
If even one NP-hard set has a (deterministic) polynomial-time
heuristic algorithm whose asymptotic error frequency is
subexponential, then the polynomial hierarchy collapses.
See the expository article of
Hemaspaandra and Williams~\shortcite{hem-wil:j:heuristic-algorithms-correctness-frequency} 
for a
discussion of that result and an attempt to reconcile it with the good
empirical results observed for hard problems.  
(Even for election problems, heuristics 
seem to often do very well, 
see, e.g.,
\cite{fal-pro:j:manipulation,rot-sch:j:typical-case-challenges}.)
Our view is that
the 
issue of proving rigorous results about the
performance of heuristics on election problems is a 
highly difficult, highly important
direction, but that NP-completeness results for a given problem are
unquestionably an excellent indication that p-time algorithms, and even p-time
heuristics with subexponential error rates, cannot be reasonably 
expected.
Thirdly, we mention that in our model, as %
is standard in this area, the
manipulators are given access to the votes of the nonmanipulators.
This is a strong though standard assumption, and admittedly is 
a model for study rather than a perfect image of the real world.  
The model actually makes the
NP-hardness results stronger (since they 
say that even with full information the problem
remains intractable) and most of our results are NP-hardness
results. 

\section{Results}
We now turn to our results regarding the complexity of 
the manipulation problem for elections, for elections with 
runoffs, and for elections with revoting runoffs.

Our results are of two basic sorts.
First, we are interested in what \emph{can} happen.
That is, for those three manipulation complexities, what is the 
relationship between them?  Is there any connection at all?  

We show that there is no connection that holds globally.  Even when limiting
ourselves just to election systems with P winner problems, we prove
that every possible case of P-or-NP-complete can simultaneously hold
for these three complexities: Each of the 
eight %
weighted and eight %
unweighted
possibilities can be realized.  The reason we want to know what
\emph{can} happen is because it is important to know the universe of
behaviors that one may face.  Note that since our runoff and revoting
problems must have the same system used in the initial and runoff
rounds, the result we mention does not follow from the important work
of 
Narodytska and Walsh~\shortcite{nar-wal:c:two-stage}
realizing all possibilities for
\xtheny{$X$}{$Y$}.%
    \footnote{%
Our complexity results regarding revoting,
which first appeared in a January 2013 technical
report on this work %
    \cite{fit-hem-hem:t:xthenx}, %
appeared after the 2012 paper of 
Narodytska and Walsh~\cite{nar-wal:c-comsoc:two-stage}
that suggested the
study of revoting but that didn't provide complexity results on
that topic; our complexity results regarding
revoting
appeared before
the 
2013 
version~\cite{nar-wal:c:two-stage} 
that has complexity results on revoting;  that is, those 
results were obtained separately.%
}

Our second type of result regards what \emph{does} happen for the most
famous, important, natural systems.  For example, although we
show that, perhaps counterintuitively, runoffs and revoting runoffs
can sometimes lower complexity and can have other bizarre relative
complexities, for none of the natural, concrete systems we have looked
at do we find this behavior to occur.  For each concrete, natural
system we have studied, runoffs and revoting runoffs either leave the
manipulation complexity unchanged, or increase the manipulation
complexity.  Of course, our results on what \emph{does} happen for 
concrete systems prove some of the cases of our 
claims regarding what \emph{can} happen.

\subsection{Realizability Theorem}\label{ss:realize}

The following theorem states our result about what can happen,
namely, regarding P and NP-completeness, any possible triple of 
complexities can occur.

\begin{theorem}\label{t:eight-part}
Let $\npccc$ denote ``$\np$-complete.''
Let $W =\allowbreak \{ (\p,\p,\p),\allowbreak (\p,\p,\npccc),\allowbreak
  (\p,\npccc,\p),\allowbreak (\p,\npccc,\npccc),\allowbreak
  (\npccc,\p,\p),\allowbreak (\npccc,\p,\npccc),\allowbreak
  (\npccc,\npccc,\p),\allowbreak (\npccc,\npccc,\npccc)\}$.
\begin{enumerate}
\item For each element $w$ of $W$, there exists an election system
$\elec$,
whose winner problem is in $\p$, such that the complexity of 
\cucm{$\elec$},
\cucmrunoff{$\elec$}, and 
\cucmrevoting{$\elec$} is, respectively, the three fields of $w$.
\item The analogous result holds for the weighted case (where the 
three fields will capture the complexity of
\cwcm{$\elec$},
\cwcmrunoff{$\elec$}, and 
\cwcmrevoting{$\elec$}, respectively).
\end{enumerate}
\end{theorem}

Recall that we promised that each election system that we construct 
to realize claims in our theorems will always
have at least one winner (if there is at least one candidate).  
The
election systems that we build in this theorem will satisfy that
also.  However, along the path to building such a system $\elec$, we will
often
first build an election system $\electwo$ that is allowed to have everyone
lose, and that satisfies the theorem 
\emph{in the model where there is always a runoff, even where there
is a unique winner in the initial round}.
We then set $\elec = \alwayswinners(\electwo)$,
where $\alwayswinners$ is as described 
in Section~\ref{sec:alwayswinners}.
Then $\elec$, which always has at least one 
winner (when there is at least one candidate),
also satisfies the conditions of the theorem 
in the model where there is always a runoff.
And since $\elec$ always has at least one winner,
$\elec$ also satisfies the conditions of the theorem in our standard 
model, where when there is a unique winner in the initial round,
this candidate is a winner of the overall election and there is
no runoff.

\def\unweightedcases{
\subsubsection{Unweighted Cases of the Realizability Theorem}\label{sss:rt}
We briefly mention
that the two most interesting unweighted cases in the realizability
theorem
are the ones 
realizing the cases 
$(\npccc,\p, \npccc)$ in Theorem~\ref{t:npc:p:npc:uw}
and, especially,
$(\npccc,\npccc,\p)$ in Theorem~\ref{t:npc:npc:p:uw}.  
The key twist in these is that both create a setting in which 
an election system can in effect pass messages to its own 
second-round self through the winner set and with the help of 
the manipulators.  In particular, in a certain set of 
circumstances, the election system can be made to, in effect, 
know that ``If the input I'm seeing is taking place in a second 
round (although I cannot myself tell whether or not it is), 
then we are utterly certainly in a model in which
revoting is allowed and indeed in which one of the manipulators has 
changed his or her vote since the initial round.''
We now present the proofs of the eight unweighted cases of
Theorem~\ref{t:eight-part}.
Most of the constructions used in these cases 
make use of the function $f(\cdot)$ described below.
\begin{definition} \label{def:fl}
Let $f(\cdot)$ denote the number of candidates needed so that
an ordering over $f(\ell)$ can express clearly at least $\ell$
possibilities. So $f(\ell)$ can be taken to be the least integer
$k$ such that $k! \geq \ell$.
\end{definition}

\begin{theorem}\label{t:npc:npc:npc:uw}
There exists an election system, $\elec$, with a p-time winner problem such
that \cucm{$\elec$}, \cucmrunoff{$\elec$}, and \cucmrevoting{$\elec$} are all
$\npc$.
\end{theorem}

\begin{proofs}
As mentioned earlier, Narodytska and Walsh~\shortcite{nar-wal:c:two-stage}
show that \cucm{STV$'$} and \cucmrunoff{STV$'$} are NP-hard, where
STV$'$ is the decisive version of STV where ties are broken in favor of the
manipulators.  As pointed out in~\cite{nar-wal:c:two-stage}, this follows
because STV$'$ = \xthenx{STV$'$}.  It is also immediate that
STV$'$ = \xthenxrevoting{STV$'$}.    This establishes
Theorem~\ref{t:npc:npc:npc:uw} for $\elec =$ STV$'$.
\end{proofs}

\begin{theorem}\label{t:p:p:p:uw}
There exists an election system, $\elec$, with a p-time winner problem such
that \cucm{$\elec$}, \cucmrunoff{$\elec$}, and \cucmrevoting{$\elec$} are all
in $\p$.
\end{theorem}

\begin{proofs}
Let $\elec = $ Plurality. Since the manipulators want to maximize the
plurality score of their preferred candidate, $p$, their optimal action
is to put $p$ at the top of their orderings. It does not benefit the
manipulators to put any other candidate higher in their orderings
or to change their vote in the runoff.
\end{proofs}

\begin{theorem}\label{t:p:npc:npc:uw}
There exists an election system, $\elec$, with a p-time winner problem such
that \cucm{$\elec$} is in $\p$ and both \cucmrunoff{$\elec$} and
\cucmrevoting{$\elec$} are $\npc$.
\end{theorem}

\begin{proofs}
Let $\elec = $ Veto. This follows directly from Theorem~\ref{t:veto},
since it is well-known that \cucm{Veto} is in $\p$.
\end{proofs}

\begin{theorem}\label{t:p:p:npc:uw}
There exists an election system, $\elec$, with a p-time winner problem such
that \cucm{$\elec$} and \cucmrunoff{$\elec$} are in $\p$, but
\cucmrevoting{$\elec$} is $\npc$.
\end{theorem}

\begin{proofs}
Let $f(\cdot)$ be as specified in Definition~\ref{def:fl}.
Let $\elec = \alwayswinners(\electwo)$, 
where $\electwo$ is defined as follows:

If $\|V\| \neq 1$ then everyone loses.\footnote{We could remove the
restriction that we have a single
voter by always allowing an arbitrary number of voters, but requiring
that all but at most one of the voters cast a vote that encodes a message
to ignore their vote.}

If $\|C\| = 2$ and the lexicographically smaller is ranked
above the lexicographically larger candidate by the voter
then everyone wins.

If $\|C\| \ge 4$, the
lexicographically smallest candidate's name encodes a formula $\psi$,
and $\|C\| \geq f(2^{\numvars(\psi)})+1$ (the ``+1'' is as we'll
not involve the lexicographically smallest candidate in encoding
the assignment),
then we do the following.
If the voter ranks the lexicographically largest above
the lexicographically smallest candidate and on all candidates other
than the lexicographically smallest it codes an assignment that
satisfies $\psi$ then the lexicographically smallest and largest
candidates win. Otherwise,
the lexicographically smallest, second smallest, and largest win.

Otherwise, everyone loses.

\noindent
That completes the specification of $\electwo$.

Now we explain why this system $\electwo$ meets all the requirements of the
theorem in the model where there is always a runoff.
\begin{itemize}
    \item Clearly the winner problem for $\electwo$ is in $\p$.
    
    \item \cucm{$\electwo$} is in $\p$ since
a preferred candidate $p$ can always be made a winner if
$\|C\| \ge 2$ and there is one manipulator and no other voters.
If $\|C\| \ge 4$, the optimal action for the manipulator is to
rank the lexicographically smallest candidate above the
lexicographically largest candidate.

    \item \cucmrunoff{$\electwo$} is in $\p$ by the following argument.
It suffices to consider the case where there is one manipulator and
no other voters.
        If the election has two candidates, the manipulator must
        rank the lexicographically smaller above the lexicographically
        larger candidate so that both candidates win the initial round
        and runoff.
If $||C|| \geq 4$, then there are zero, two,
 or three winners in the initial round.
If there are three winners in the initial round, there are no winners
in the runoff.  If there are two winners in the initial round, 
then the lexicographically largest candidate is ranked above
the lexicographically smallest candidate.  But then there are also
no winners in the runoff without revoting.

    \item \cucmrevoting{$\electwo$} is $\npc$ by reducing from $\sat$.   
        Observe that $\psi \in \sat$ reduces to \cucmrevoting{$\electwo$} for
        the candidate set: $p$ encoding $\psi$ and
        $\text{max}(3,f(2^{\numvars(\psi)}))$ dummy candidates all
        lexicographically larger than $p$. Let the preferred candidate be
        $p$ and let there be zero nonmanipulative voters and one manipulative
        voter.  In order for $p$ to win the runoff, there have to be
two winners in the initial round.   This implies that $\psi$ is satisfiable.
And if $\psi$ is satisfiable, $p$ can be made a winner by
having the manipulator vote so that
$p$ is ranked last and
        its ordering over the dummy candidates codes a satisfying
        assignment to $\psi$.
        In the runoff the manipulator must change her
        vote so that $p$ is ranked first.
\end{itemize}
So $\electwo$ satisfies the theorem in
the model where there is always a runoff, and
$\alwayswinners(\electwo)$ satisfies the theorem in our standard model.~\end{proofs}

\begin{theorem}\label{t:npc:p:npc:uw}
There exists an election system, $\elec$, with a p-time winner problem such
that \cucm{$\elec$} is $\npc$, \cucmrunoff{$\elec$} is in $\p$, and
\cucmrevoting{$\elec$} is $\npc$.
\end{theorem}

\begin{proofs}
Let $f(\cdot)$ be as specified in Definition~\ref{def:fl}.
Let $\elec = \alwayswinners(\electwo)$, 
where $\electwo$ is defined as follows:
If $\|V\| \neq 1$ or $\|C\| = 1$ then everyone loses.
Otherwise, we'll handle things based on the number of candidates and the single
voter $v$, as described below.

If $\|C\| = 2$ and the lexicographically larger
candidate is more preferred
than the lexicographically smaller
candidate by the voter %
then everyone wins,
else everyone loses.

If $\|C\| \geq 3$, the lexicographically smallest candidate's name
encodes a formula $\psi$, $\|C\| \ge f(2^{\numvars(\psi)})+1$
(the ``$+1$'' is as we'll not involve the lexicographically smallest
candidate in encoding the assignment),
$v$'s
vote on all candidates
other than the lexicographically smallest candidate codes an assignment
that satisfies $\psi$, and
the lexicographically smallest candidate is more preferred than the
lexicographically largest candidate by $v$, then
the lexicographically largest and smallest candidates win, else
everyone loses.

\noindent
That completes the specification of $\electwo$.

Now we explain why this system $\electwo$ meets all the requirements of the
theorem in the model where there is always a runoff.
\begin{itemize}
    \item Clearly the winner problem for $\electwo$ is in $\p$.
  
    \item \cucm{$\electwo$} is $\npc$ by reducing from $\sat$. Observe
      that $\psi \in \sat$ reduces to \cucm{$\electwo$} for the candidate set:
      $p$ encoding $\psi$ and $\max(2, f(2^{\numvars(\psi)}))$ dummy candidates
      all lexicographically larger than $p$.
      Let the preferred candidate be $p$ and let there
      be zero nonmanipulative voters and one manipulative voter. The manipulator
      must vote with $p$ at the top of her preference order and code a
      satisfying assignment to $\psi$ over the dummy candidates.
      If $\psi \not \in \sat$ then no one wins. Therefore \cucm{$\electwo$} is
      $\npc$.

    \item \cucmrunoff{$\electwo$} is in $\p$ since 
      in the initial %
      round we always have zero or two winners.
      However, when we have two winners (and more than two candidates),
      they always lose in the second
      round unless the manipulator %
      can change her vote. %
      So in the second round scheme
      (runoff without revoting), no one ever can win if we have
      more than two candidates.
   
    \item \cucmrevoting{$\electwo$} is $\npc$ by the
     same argument that we used to show that \cucm{$\electwo$} is $\npc$.
     Here, it works by, when the $\psi$ encoded by the lexicographically
     smallest candidate 
     is satisfiable, the manipulator in the initial %
     round conveys
     (in the dummy candidates)
     a satisfying assignment, and she %
     puts the lexicographically smallest 
     candidate as her most preferred candidate.
     In the runoff %
     the manipulator puts the lexicographically smallest
     candidate as her least preferred candidate.
     (So we'll go to the $\|C\| = 2$ case 
     of $\electwo$ in the runoff %
     and we will have two %
     winners, including the lexicographically smallest candidate.)
\end{itemize}
So $\electwo$ satisfies the theorem in
the model where there is always a runoff, and
$\alwayswinners(\electwo)$ satisfies the theorem in our standard model.
\end{proofs}

\begin{theorem}\label{t:npc:p:p:uw}
There exists an election system, $\elec$, with a p-time winner problem such
that \cucm{$\elec$} is $\npc$, but \cucmrunoff{$\elec$} and
\cucmrevoting{$\elec$} are in $\p$.
\end{theorem}

\begin{proofs}
Let $\elec = \alwayswinners(\electwo)$, 
where $\electwo$ is defined as follows:

If $\|C\| \geq 2$ then STV$'$ (where STV$'$ is as defined in
Theorem~\ref{t:npc:npc:npc:uw}). Otherwise, everyone loses.

That completes the specification of $\electwo$.

Now we explain why this system $\electwo$ meets all the requirements of the
theorem in the model where there is always a runoff.
\begin{itemize}
    \item Clearly $\electwo$ has a p-time winner problem.

    \item \cwcm{$\electwo$} is $\npc$ since
 \cucm{$\electwo$} on at least two candidates 
corresponds to \cucm{STV$'$}.

    \item \cucmrunoff{$\electwo$} and \cucmrevoting{$\electwo$} are
        each in $\p$ since the initial round results in at most one winner,
        and so the runoff will never give a winner (even with revoting).
\end{itemize}
So $\electwo$ satisfies the theorem in
the model where there is always a runoff, and
$\alwayswinners(\electwo)$ satisfies the theorem in our standard model.~\end{proofs}
\begin{theorem}\label{t:p:npc:p:uw}
There exists an election system, $\elec$, with a p-time winner problem such
that \cucm{$\elec$} is in $\p$, \cucmrunoff{$\elec$} is $\npc$, and
\cucmrevoting{$\elec$} is in $\p$.
\end{theorem}

\begin{proofs}
Let $\elec = \alwayswinners(\electwo)$, 
where $\electwo$ is defined as follows.

We will utilize several different special candidate names in our proof.
The candidate names are as follows:
\begin{itemize}
    \item $\pair{\text{Shiva}_1,\epsilon}$, which we refer to as Shiva$_1$.
    \item $\pair{\text{Shiva}_2,\psi}$, where $\psi$ is a boolean formula,
          which we refer to as Shiva$_2$-like.
    \item $\pair{\text{Angel},1}$, which we refer to as Angel$_1$.
    \item $\pair{\text{Angel},2}$, which we refer to as Angel$_2$.
    \item More angels are used as needed.
\end{itemize}
Let $f(\cdot)$ be as specified in Definition~\ref{def:fl}.
For our election $\electwo$, the candidate set is expected to be
one of the following forms:
\begin{itemize}
    \item[(a)] One Shiva$_2$-like candidate with formula $\psi$, the Shiva$_1$
    	candidate, and enough angel candidates to encode $2^{\numvars(\psi)}+1$
        possibilities ($2^{\numvars(\psi)}$ assignments to $\psi$ and
one special ``Begone-2'' ordering),
i.e., $f(2^{\numvars(\psi)}+1)$ angel candidates.
        
    \item[(b)] One Shiva$_2$-like candidate with formula $\psi$
        and enough angel candidates to encode $2^{\numvars(\psi)}+1$
        possibilities, i.e., $f(2^{\numvars(\psi)}+1)$ angel candidates.
\end{itemize}
Let $\electwo$ be defined as follows:

If $\|V\| \neq 1$ or $C$ is not of an expected form then everyone loses.

If $C$ is of form (a) and the angel candidates' restriction of the voter's vote
encodes an assignment to $\psi$, then the Shiva$_2$-like candidate and all
of the angel candidates win, else everyone loses.

If $C$ is of form (b) and the angel candidates' restriction of the voter's vote
encodes the special ``Begone-2'' ordering or a satisfying assignment to $\psi$
then all of the angels win, else everyone loses.

\noindent
That completes the specification of $\electwo$.

Now we explain why this system $\electwo$ meets all the requirements of the
theorem in the model where there is always a runoff.
\begin{itemize}
    \item Clearly the winner problem for $\electwo$ is in $\p$.
    
    \item \cucm{$\electwo$} is in $\p$ by looking at both of the allowed
        candidate set forms:
    \begin{itemize}
        \item For form (a), when the voter casts a vote that encodes an
            assignment to $\psi$ over the angels then everyone wins
            who can possibly win.
        \item For form (b), when the voter casts a vote that encodes the
            special ``Begone-2'' ordering over the angels then everyone
            wins who can possibly win.
    \end{itemize}

    \item \cucmrunoff{$\electwo$} is $\npc$ by reducing from $\sat$. Observe that
    to see if $\psi \in \sat$ ask if Angel$_1$ can win the
    \cucmrunoff{$\electwo$} instance with the candidate set:
    Shiva$_1$, 
    $\pair{\text{Shiva}_2,\psi}$, and $f(2^{\numvars(\psi)}+1)$ angels.
    Let the voter set contain zero nonmanipulative voters and
    one manipulative voter.
    
    If $\psi \in \sat$, then the manipulator casts the initial-round
    vote that corresponds to a satisfying assignment to $\psi$,
    so the Shiva$_2$-like candidate and all of the angels win in
    the initial round and all of the angels win in the runoff.
    Conversely, if $\psi \notin \sat$ then Angel$_1$ would not be able
    to win in the runoff.
    
    \item \cucmrevoting{$\electwo$} is in $\p$ by looking at both of the
        allowed candidate set forms:
    \begin{itemize}
        \item For form (a), when the voter casts a vote that codes an
            assignment to $\psi$ over the angels in the initial round
            and then changes her vote to encode the special ``Begone-2''
            ordering over the angels then everyone wins who can
            possibly win.
        \item For form (b), everyone loses by the end of the runoff.
    \end{itemize}
\end{itemize}
So $\electwo$ satisfies the theorem in
the model where there is always a runoff, and
$\alwayswinners(\electwo)$ satisfies the theorem in our standard model.
\end{proofs}

\begin{theorem}\label{t:npc:npc:p:uw}
There exists an election system, $\elec$, with a p-time winner problem such
that \cucm{$\elec$} and \cucmrunoff{$\elec$} are $\npc$, but
\cucmrevoting{$\elec$} is in $\p$.
\end{theorem}

\begin{proofs}
Let $f(\cdot)$ be as specified in Definition~\ref{def:fl}.
Let $\elec = \alwayswinners(\electwo)$, 
where $\electwo$ is defined as follows.
The allowed (all others will cause everyone to lose) candidate types for
$\electwo$ are described in Table~\ref{allow:c:type}.

\begin{table}[t]
\centering
\def\arraystretch{1.5}
\begin{tabular}{  l | p{'220pt} } 
    \text{Candidate Form} & \text{Role in our proof} \\ \hline
    $\pair{1,1}$ & Seeks to make the initial round hard. \\ %
    $\pair{1,2}$ & Seeks to make the initial round easy. \\ %
    $\pair{2,\psi}$ & Candidate coding a formula intended as
                 part of a hard initial round. \\ %
    $\pair{3,\psi}$ & Candidate coding a formula intended as part
                of a hard runoff.\\ %
    $\pair{4,\mbox{(any string)}}$ & ``Type-4'' dummy candidate, 
                    used to make votes so big as to encode assignments. \\ 
    $\pair{5,1},\pair{5,2}$ & Special dummy candidates to allow vote changes 
                to show through in some cases. \\ 
\end{tabular}
    \caption{{}Allowed candidate types for the election $\electwo$ used in the
    proof of Theorem~\ref{t:npc:npc:p:uw}\label{allow:c:type}.}
\end{table}

For our election $\electwo$, the candidate set is expected to be in one of the
following forms (we assume w.l.o.g.\ that every formula has at least one
variable):
\begin{itemize}
    \item[(a)] $\pair{1,1},\pair{2,\psi}$, and enough
        type-4 dummy candidates so that a vote including
        them can encode an assignment to $\psi$, i.e., at
        least $f(2^{{\numvars(\psi)}})$ dummy candidates.
    \item[(b)] $\pair{1,2},\pair{3,\psi},\pair{5,1},\pair{5,2}$, and enough
        type-4 dummy candidates so that a vote including
        them can encode an assignment to $\psi$, i.e., at
        least $f(2^{{\numvars(\psi)}})$ dummy candidates.
    \item[(c)] $\pair{3,\psi},\pair{5,1},\pair{5,2}$, and enough
        type-4 dummy candidates so that a vote including
        them can encode an assignment to $\psi$, i.e., at
        least $f(2^{{\numvars(\psi)}})$ dummy candidates.
\end{itemize}
Let $\electwo$ be defined as follows:

If $\|V\| \neq 1$ or $C$ is not of an expected form
then everyone loses.
Otherwise, we'll handle things as 
described below (note that in this case there is a single voter $v$).

If $C$ is of form (a) and the type-4 dummy candidates'
restriction of $v$'s vote encodes a satisfying assignment
to $\psi$, then $\pair{2,\psi}$ wins.
Otherwise, everyone loses.

If $C$ is of form (b) and $\pair{5,2} > \pair{5,1}$ in $v$'s vote
then $\pair{3,\psi},\pair{5,1},\pair{5,2},$ and all of the type-4 dummy
candidates win.
In all other cases where $C$ is of form (b), everyone loses.
        
If $C$ is of form (c) and $\pair{5,1} > \pair{5,2}$ in $v$'s vote
or the type-4 dummy candidates' restriction of $v$'s vote encodes
a satisfying assignment to $\psi$, then $\pair{3,\psi}$ wins.
Otherwise, everyone loses.

\noindent
This completes the specification of $\electwo$.

Now we must ensure that $\electwo$ meets all of the requirements of this
theorem in the model where there is always a runoff.

\begin{itemize}
    \item Clearly the winner problem for $\electwo$ is in $\p$.
    \item \cucm{$\electwo$} is $\npc$ since to test if $\psi
         \in \sat$, we can ask if $\pair{2,\psi}$ can win the
         \cucm{$\electwo$} instance
          with
          candidate set \{$\pair{1,1},\pair{2,\psi},f(2^{{\numvars(\psi)}})$
          type-4 dummy candidates\} and the voter set containing
          zero nonmanipulative voters and one manipulative voter.
          For $\pair{2,\psi}$ to win the manipulator must code a satisfying
          assignment to $\psi$ in her
          ordering over the type-4 dummy
          candidates. If $\psi \notin \sat$, then no one can win.

    \item \cucmrunoff{$\electwo$} is $\npc$ since to test
          if $\psi \in \sat$, we can ask if $\pair{3,\psi}$ can win the
          \cucmrunoff{$\electwo$} instance
          with candidate set \{$\pair{1,2},\pair{3,\psi},\pair{5,1},\pair{5,2},
          f(2^{\numvars(\psi)})$ type-4
          dummy candidates\} and voter set containing
          zero nonmanipulative voters and one manipulative voter.

          Note that we need $\pair{5,2} > \pair{5,1}$ in the manipulator's
          vote in order for $\pair{3,\psi}$ to make it to the runoff.
          If $\psi \notin \sat$, then $\pair{3,\psi}$ cannot win. Otherwise,
          if $\psi \in \sat$, voting so that $\pair{5,2} > \pair{5,1}$ in the
          manipulator's vote and with the type-4 dummy order giving a
          satisfying assignment to $\psi$, makes $\pair{3,\psi}$ a winner.
    \item \cucmrevoting{$\electwo$} is in $\p$ by looking at each of the allowed
        candidate set forms:\hfill
        \begin{itemize}
            \item For form (a),
everyone loses by the end of the runoff.
            \item For form (b), when voter $v$ casts a vote where
            $\pair{5,2} > \pair{5,1}$
            in the initial %
            round and a vote where $\pair{5,1} > \pair{5,2}$
            in the runoff, %
            then everyone wins who can possibly win.
            \item For form (c), everyone loses by the end of the runoff.
        \end{itemize}
        All other cases have everyone lose immediately.
        Thus \cucmrevoting{$\electwo$} is in $\p$.
\end{itemize}
So $\electwo$ satisfies the theorem in
the model where there is always a runoff, and
$\alwayswinners(\electwo)$ satisfies the theorem in our standard model.
\end{proofs}
} %

\def\weightedcases{
\subsubsection{Weighted Cases of the Realizability Theorem}
\label{sss:weighted}
We now present the proofs of the eight weighted cases of
Theorem~\ref{t:eight-part}.
As is typical for NP-hardness proofs for weighted manipulation, we will
usually
reduce from the well-known NP-complete problem Partition: Given a 
nonempty set of positive integers $k_1, \ldots, k_t$ that sums to $2K$,
we ask if $k_1, \ldots, k_t$ can be partitioned into two subsets of
equal size.
\begin{theorem}\label{t:npc:npc:npc:w}
There exists an election system, $\elec$, with a p-time winner problem such
that \cwcm{$\elec$}, \cwcmrunoff{$\elec$}, and \cwcmrevoting{$\elec$} are all
$\npc$.
\end{theorem}

\begin{proofs}
Let $\elec$ be Veto.
This follows directly from~\cite{hem-hem:j:dichotomy}
and Theorem~\ref{t:plur:triv}.~\end{proofs}

\begin{theorem}\label{t:p:p:p:w}
There exists an election system, $\elec$, with a p-time winner problem such
that \cwcm{$\elec$}, \cwcmrunoff{$\elec$}, and \cwcmrevoting{$\elec$} are all
in $\p$.
\end{theorem}

\begin{proofs}
Let $\elec$ be Plurality. This follows directly from
Theorem~\ref{t:plurality}.~\end{proofs}

\begin{theorem}\label{t:p:npc:npc:w}
There exists an election system, $\elec$, with a p-time winner problem such
that \cwcm{$\elec$} is in $\p$ and both \cwcmrunoff{$\elec$} and
\cwcmrevoting{$\elec$} are $\npc$.
\end{theorem}

\begin{proofs}
Let $\elec$ be defined as follows:

If $\|C\| \leq 4$ then Llull, else everyone wins.

\noindent
That completes the specification of $\elec$.

This follows directly from~\cite{fal-hem-sch:c:llull4} and
Theorem~\ref{t:llull}.~\end{proofs}
\begin{theorem}\label{t:p:p:npc:w}
There exists an election system, $\elec$, with a p-time winner problem such
that \cwcm{$\elec$} and \cwcmrunoff{$\elec$} are in $\p$, but
\cwcmrevoting{$\elec$} is $\npc$.
\end{theorem}

\begin{proofs}
Let $\elec = \alwayswinners(\electwo)$, 
where $\electwo$ is defined as follows:

If $\|C\| = 3$ then candidates with plurality scores $\geq 50\%$ win.

If $\|C\| = 2$ then candidates with plurality scores
$> 50\%$ win.

Otherwise, everyone loses.

\noindent
That completes the specification of $\electwo$.

Now we explain why this system $\electwo$ meets all the requirements of the
theorem in the model where there is always a runoff.
\begin{itemize}
    \item Clearly the winner problem for $\electwo$ is in $\p$.
    \item \cwcm{$\electwo$} is in $\p$ since in all cases where a
        preferred candidate can win, the optimal action for the manipulators
        is to vote for that candidate.
    \item \cwcmrunoff{$\electwo$} is in $\p$ since the only way to have winners
        in a two-stage $\electwo$ election is to have three candidates in the
        initial round and two candidates in the runoff. No candidates can
        win in the runoff unless revoting is allowed, so
        \cwcmrunoff{$\electwo$} is in $\p$.
    \item \cwcmrevoting{$\electwo$} is $\npc$ since to test if a
        set of positive integers
        $k_1, \ldots, k_t \in $ Partition
        we can ask if the candidate $p$ can win
        the \cwcmrevoting{$\electwo$} instance with the candidate
        set $\{p,r,\ell\}$ and the voter
        set containing zero nonmanipulative voters and $t$ manipulative voters.
        Let the $t$ manipulators have weights that correspond to
        the Partition instance,
        i.e., $k_1, \ldots, k_t$ such that $\sum_{i=1}^{t} k_i = 2K$.

        To ensure that $p$ is an overall winner, the manipulators must
        partition %
        their votes between $p$ and w.l.o.g.\ $r$,
        since the only way to have
        a runoff where a candidate can win is when two candidates win
        in the initial %
        round. Then in the runoff the manipulators all
        vote for $p$. Thus if there is a way for the manipulators
        to partition their votes
        into two equal-weight subsets, then $p$
        can be made the overall winner. Otherwise, $p$ cannot be made
        the overall winner.
        Therefore \cwcmrevoting{$\electwo$} is $\npc$.
\end{itemize}
So $\electwo$ satisfies the theorem in
the model where there is always a runoff, and
$\alwayswinners(\electwo)$ satisfies the theorem in our standard model.
~\end{proofs}

\begin{theorem}\label{t:npc:p:npc:w}
There exists an election system, $\elec$, with a p-time winner problem such
that \cwcm{$\elec$} is $\npc$, \cwcmrunoff{$\elec$} is in $\p$, and
\cwcmrevoting{$\elec$} is $\npc$.
\end{theorem}

\begin{proofs}
Let $\elec = \alwayswinners(\electwo)$, 
where $\electwo$ is defined as follows:

If $\|C\|=3$ then candidates with plurality scores
of exactly $50\%$ win.

If $\|C\|=2$ then candidates with plurality scores $> 50\%$ win.

Otherwise, everyone loses.

\noindent
That completes the specification of $\electwo$.

Now we explain why this system $\electwo$ meets all the requirements of the
theorem in the model where there is always a runoff.
\begin{itemize}
    \item Clearly the winner problem for $\electwo$ is in $\p$.
    
    \item \cwcm{$\electwo$} is $\npc$ since to test if a set of
        positive integers $k_1, \ldots, k_t \in $ Partition
        we can ask if the candidate $p$ can win
        the \cwcm{$\electwo$} instance with the candidate set $\{p,r,\ell\}$ and
        the voter
        set containing zero nonmanipulative voters and $t$ manipulative voters.
        Let the $t$ manipulators have weights that correspond
        to the Partition instance,
        i.e., $k_1, \ldots, k_t$ such that $\sum_{i=1}^{t} k_i = 2K$.

Since $\|C\|=3$, $p$ is a winner if and only if $p$ has a score
of exactly 50\% of the available vote weight.  So, $p$ can be made a winner
if and only of the manipulators can partition their votes into two
equal-weight subsets.

    \item \cwcmrunoff{$\electwo$} is in $\p$ since no one can win in the runoff
        without revoting. If the initial round has three candidates where
        two attain exactly 50\% of the vote to move on to the runoff, no
        one will win in the runoff since neither will have greater than
        $50\%$ of the
        available vote weight. If the initial round has only two candidates,
        the winner will lose the runoff since when there is only one candidate,
        that candidate always loses.
        In all other cases the candidates all lose immediately.
        Therefore \cwcmrunoff{$\electwo$} is in $\p$.
    
    \item \cwcmrevoting{$\electwo$} is $\npc$ since to test if a set of
        positive integers $k_1, \ldots, k_t \in $ Partition
        we can ask if the candidate $p$ can win
        the \cwcmrevoting{$\electwo$} instance with the
        candidate set $\{p,r,\ell\}$ and the voter
        set containing zero nonmanipulative voters and $t$ manipulative voters.
        Let the $t$ manipulators have weights that correspond
        to the Partition instance,
        i.e., $k_1, \ldots, k_t$ such that $\sum_{i=1}^{t} k_i = 2K$.
        
        To ensure that $p$ is an overall winner, the manipulators must
        partition %
        their votes between $p$ and w.l.o.g.\ $r$,
        since the only way to have
        a runoff where a candidate can win is when two candidates win
        in the initial round. Then in the runoff the manipulators all
        vote for $p$ to increase $p$'s score to be greater than $50\%$.
        Thus if there is a way to partition the vote
        weight of the manipulators into two equal-weight subsets, then $p$
        can be made the overall winner. Otherwise, $p$ cannot be made
        the overall winner. Therefore \cwcmrevoting{$\electwo$} is $\npc$.
\end{itemize}
So $\electwo$ satisfies the theorem in
the model where there is always a runoff, and
$\alwayswinners(\electwo)$ satisfies the theorem in our standard model.~\end{proofs}
\begin{theorem}\label{t:npc:p:p:w}
There exists an election system, $\elec$, with a p-time winner problem such
that \cwcm{$\elec$} is $\npc$, but \cwcmrunoff{$\elec$} and
\cwcmrevoting{$\elec$} are in $\p$.
\end{theorem}

\begin{proofs}
Let $\elec = \alwayswinners(\electwo)$, 
where $\electwo$ is defined as follows:

If $\|C\| \geq 2$ then the candidate with the highest veto score
wins as long as the winner is unique. Otherwise (if there is no unique
winner or $\|C\| = 1$), everyone loses.

\noindent
That completes the specification of $\electwo$

Now we explain why this system $\electwo$ meets all the requirements of the
theorem in the model where there is always a runoff.
\begin{itemize}
    \item Clearly $\electwo$ has a p-time winner problem.

    \item \cwcm{$\electwo$} is $\npc$ since
 \cwcm{$\electwo$} restricted to three candidates
corresponds to \cwcm{Veto} restricted to three candidates in the
unique winner model, which is known to be
$\npc$~\cite{con-lan-san:j:when-hard-to-manipulate}.

    \item \cwcmrunoff{$\electwo$} and \cwcmrevoting{$\electwo$} are
        each in $\p$ since the initial round results in at most one winner,
        and so the runoff will never give a winner (even with revoting).
\end{itemize}
So $\electwo$ satisfies the theorem in
the model where there is always a runoff, and
$\alwayswinners(\electwo)$ satisfies the theorem in our standard model.~\end{proofs}
\begin{theorem}\label{t:p:npc:p:w}
There exists an election system, $\elec$, with a p-time winner problem such
that \cwcm{$\elec$} is in $\p$, \cwcmrunoff{$\elec$} is $\npc$, and
\cwcmrevoting{$\elec$} is in $\p$.
\end{theorem}

\begin{proofs}
Let $\elec = \alwayswinners(\electwo)$, 
where $\electwo$ is defined as follows:

If $\|C\| = 3$ and there is a candidate with plurality score 0 
then candidates
with plurality scores greater than or equal to half of the
highest occurring score win.

If $\|C\| = 2$ then candidates with plurality scores
less than or equal to half of the highest occurring plurality score win.

If $\|C\| = 1$ then that candidate wins.

Otherwise, everyone loses.

\noindent
That completes the specification of $\electwo$.

Now we explain why this system $\electwo$ meets all the requirements of the
theorem in the model where there is always a runoff.
\begin{itemize}
    \item Clearly $\electwo$ has a p-time winner problem.

    \item \cwcm{$\electwo$} is in $\p$ since the optimal action for the
        manipulators is to vote for their preferred candidate
        $p$ (when $\|C\| = 3$ or $\|C\|=1$) to maximize $p$'s score or to
        not vote for $p$ (when $\|C\|=2$) to minimize $p$'s score.

    \item \cwcmrunoff{$\electwo$} is $\npc$ since to test if a set
        of positive integers
        $k_1, \ldots, k_t \in $ Partition we can ask if
        the candidate $p$ can win
        the \cwcmrunoff{$\electwo$} instance with
        the candidate set $\{p,r,\ell\}$ and the voter
        set containing one nonmanipulative voter and $t$ manipulative voters.
        Let the nonmanipulator have weight $K$ and vote
        $r > p > \ell$ and let the $t$ manipulative voters have weights that
        correspond to the Partition instance, i.e., $k_1, \ldots, k_t$
        such that $\sum_{i=1}^{t} k_i = 2K$.
        
        To ensure that the preferred candidate $p$ wins,
$\ell$ needs to score 0 and
        $p$ needs to score at least
        $K$ to advance to the runoff.
Then $p$ and $r$ will proceed to the runoff.  In order for $p$ to win
the runoff, $p$ needs to score at most $K$.  So, $p$ needs to score
exactly $K$ in order to become a winner. 

    \item \cwcmrevoting{$\electwo$} is in $\p$ by the following argument.
If $V = \emptyset$, simply check if $p$ is a winner of the two-stage election.
Now assume that $V \neq \emptyset$.  If $\|C\| = 1$, $p$ will
always be a winner.  If $\|C\| > 3$, there are no winners.
If $\|C\| = 2$, the runoff will consist of 
at most one candidate.  So, the optimal action is for all manipulators
to not vote for $p$ in the initial round, to maximize $p$'s chances of
participating in the runoff.
Finally, let $\|C\| = 3$.
If $p$ can be made the unique winner of the initial round,
then $p$ wins the overall election.  If that is not possible, 
and $p$ makes it to the runoff, there are two participants in the runoff.
In both cases, the optimal action for the manipulators in the initial
round is to vote for $p$.  And if the runoff has two participants, the
optimal action in the runoff is to not vote for $p$.

\end{itemize}
So $\electwo$ satisfies the theorem in
the model where there is always a runoff, and
$\alwayswinners(\electwo)$ satisfies the theorem in our standard model.~\end{proofs}

We add an additional case to the election system used in the proof of the
previous case to raise the complexity of manipulation to
be $\npc$, while keeping the complexity of manipulation with
revoting runoffs in $\p$.
\begin{theorem}\label{t:npc:npc:p:w}
There exists an election system, $\elec$, with a p-time winner problem such
that \cwcm{$\elec$} and \cwcmrunoff{$\elec$} are $\npc$, but
\cwcmrevoting{$\elec$} is in $\p$.
\end{theorem}

\begin{proofs}
Let $\elec = \alwayswinners(\electwo)$, 
where $\electwo$ is defined as follows:

If $\|C\| = 5$ and there are exactly four candidates that
have plurality scores of exactly $25\%$, then these four candidates win.

If $\|C\| = 3$ and there is a candidate with plurality score 0 
then candidates
with plurality scores greater than or equal to half of the
highest occurring score win.

If $\|C\| = 2$ then candidates with plurality scores less than
or equal to half of the highest occurring plurality score win.

If $\|C\| = 1$ then that candidate wins.

Otherwise, everyone loses.

\noindent
That completes the specification of $\electwo$.

Now we explain why this system $\electwo$ meets all the requirements of the
theorem in the model where there is always a runoff.
\begin{itemize}
    
    \item Clearly $\electwo$ has a p-time winner problem.
    
    \item \cwcm{$\electwo$} is $\npc$ since to test if a set
        of positive integers $k_1, \ldots, k_t \in $ Partition
        we can ask if the candidate $p$ can win
        the \cwcm{$\electwo$} instance with the
        candidate set $C = \{a,b,c,p,\ell\}$,
one weight-$K$ nonmanipulator voting for $a$, 
one weight-$K$ nonmanipulator voting for $b$, 
and $t$ manipulative voters.
         Let the $t$ manipulative voters have weights that correspond to the
         Partition instance, i.e., $k_1, \ldots, k_t$ such that
         $\sum k_i = 2K$.
It is immediate that $p$ is a winner if and only if exactly half of the
manipulator weight votes for $p$.  This is possible if and only if
$k_1, \ldots, k_t \in$ Partition.

    \item \cwcmrunoff{$\electwo$} is $\npc$ by exactly the same argument
as in the proof of Theorem~\ref{t:p:npc:p:w}.

    \item \cwcmrevoting{$\electwo$} is in $\p$ by the same
        argument as used above for the previous case
        (Theorem~\ref{t:p:npc:p:w}).
Observe that the addition of
        the $\|C\|=5$ case does not increase the complexity of
        \cwcmrevoting{$\electwo$}:
There are never five candidates in the runoff and 
if there are five candidates in the initial round,
there will be zero or four candidates in the runoff,
and so there will be no winners in the runoff    

\end{itemize}
So $\electwo$ satisfies the theorem in
the model where there is always a runoff, and
$\alwayswinners(\electwo)$ satisfies the theorem in our standard model.~\end{proofs}

}%

\unweightedcases

\weightedcases

\def\alwayswinnersconstruction{
\subsection{\boldmath The $\alwayswinners$ Construction}
\label{sec:alwayswinners}

We now describe the transformation $\alwayswinners$ that was used in the proofs
of many of the cases of Theorem~\ref{t:eight-part}.
What this transformation will do is 
it will take an election system and will transform it into a new 
election system that will (except when the candidate 
set is empty) always have at least 
one winner, yet that is so closely related to the original election system
that the complexity of the original system is unchanged, 
with respect to the three problems we are concerned with here
\emph{in the model where there is always a runoff, even where there
is a unique winner in the initial round} (and in fact,
with respect to many, many other types of manipulative actions).
Note that once we have an election system that will
always have at least one winner, the model in which there is always 
a runoff coincides with our standard model.

This transformation thus relatively broadly addresses a persistently 
annoying issue.  Many people, including the authors, feel that it is 
ugly and asymmetric to require nonempty winner sets yet to allow 
all of $C$ to be the winner set.  And indeed a number of papers in 
\emph{computational} social choice 
do allow the winner set to be 
any member of $2^C$.  On the other hand, traditionally in 
\emph{social choice}, the requirement that the winner set be nonempty
is part of the definition of elections.  Our transformation 
shows that the difference in models isn't as large as one might 
think; one can often adopt the more symmetric model, yet by this 
transformation one will know that one's results also hold in the 
more restrictive model.  

We now give the transformation $\alwayswinners$.  First let us describe
informally how it works, and then we'll describe it and its properties
more formally.  Let $\elece$ be the election system we want to
transform under $\alwayswinners$.  Informally, suppose that we have a
candidate name, $new$, that is not part of our universe of legal
candidate names.  (Of course this is untrue; the universe of names is
as it is.  But please indulge us for a few more lines, and we'll then 
avoid this problem 
in our more formal approach.)  
Then under $\alwayswinners$, if $new$ is not an input
candidate then all candidates win, and if $new$ is an input candidate
then $new$ wins and also all candidates win who under $\elece$ would
win if in our input election $new$ is removed and all votes are masked
down to remove $new$.  Note that this system always has a winner (if 
the set of candidates is not empty).  Also, it can be easily 
seen to retain
most of the manipulative-action 
complexities of $\elece$, as we'll discuss later.

As admitted above, we can't just make a new name appear.  But we can get 
the same effect formally, by shifting all the names in the universe up 
by one spot to open up a space for our new name, and then when we're 
using those other names in simulating the original system, by shifting 
them back
down again.  And that is precisely what we will do.  

So we now more carefully and correctly specify the transformation
$\alwayswinners$.
\begin{definition} \label{def:alwayswinners}
Let $\elece$ be 
an election system (that perhaps has no winners even on inputs
on which $C$ is nonempty).  Let the set of legal names for 
candidates (i.e., the set of all strings) be enumerated in 
lexicographic order by $s_0, s_1, s_2, \ldots$ (e.g.,
``$\epsilon$, 0, 1, 00, 01,~$\ldots$'' if names are taken to be 
binary strings).  Let $\texttt{++}$
denote a one-step increase in this order, i.e., $s_i{\texttt{++}} = s_{i+1}$.
The $\texttt{++}$ operator naturally applies to sets of candidates, namely
as defined by 
$A{\texttt{++}} = \{ a\texttt{++} \condition a \in A\}$.  And for any set 
$A$ of candidates such that $s_0 \not\in A$, we similarly 
define the decrement of the set, namely by
$A\mbox{-\,-} = \{ a \condition a{\texttt{++}} \in A\}$.
On candidate set $C$ and voter set $V$, $\alwayswinners(\elece)$ does 
the following.  If $s_0 \not\in C$, then the winner set is $C$.
If $s_0 \in C$ then the winner set is $
\elece( 
(C - \{ s_0  \})\mbox{-\,-}, V')\texttt{++} \cup 
\{s_0\}$, 
where $V'$ is $V$ with $s_0$ masked 
out of each preference order and then each candidate name decremented
in each order and where $\elece(\hat{C},\hat{V})$ denotes the winner 
set, under $\elece$, of the election over candidate set $\hat{C}$
and voter set $\hat{V}$.
\end{definition}

The crucial things to notice about $\alwayswinners(\elece)$ are 
the following, which hold for all election systems $\elece$ (including
ones that allow there to be no winners on some inputs for which
the input candidate set of the 
$\elece$ instance is nonempty)
\emph{in the model 
where there is always a runoff}.
$\alwayswinners(\elece)$ always has at least one 
winner 
(when the input candidate set of the 
$\alwayswinners(\elece)$ instance is nonempty).  
For \cucm{$\elece$} (respectively, \cucmrunoff{$\elece$},
\cucmrevoting{$\elece$}) it holds that if the problem is in $\p$ 
then \cucm{$\alwayswinners(\elece)$} (respectively,
\cucmrunoff{$\alwayswinners(\elece)$},
\cucmrevoting{$\alwayswinners(\elece)$}) is in $\p$.
For \cucm{$\elece$} (respectively, \cucmrunoff{$\elece$},
\cucmrevoting{$\elece$}) it holds that if the problem is $\npc$ 
then \cucm{$\alwayswinners(\elece)$} (respectively,
\cucmrunoff{$\alwayswinners(\elece)$},
\cucmrevoting{$\alwayswinners(\elece)$}) is $\npc$.
Part of the easy task 
of seeing that these complexity connections hold is noticing 
that given an instance of one of these problems under $\elece$,
one can increment all candidate names both within the 
candidate set and the voter preferences, can then add in 
a new candidate $s_0$ and extend voter preferences arbitrarily 
to include that new candidate (e.g., putting it last in each 
voter's preferences),  and then we can note that a candidate $p$ 
can be made a winner 
in the initial election under $\elece$ exactly if 
$p\texttt{++}$ can be made a winner 
under $\alwayswinners(\elece)$ in the transformed election.  (And 
we mention in passing that the 
analogous claim holds for the so-called ``destructive'' case in 
which we seek to preclude $p$ from being a winner, though destructive 
cases are not a focus of this paper.)

The observations above are what we need to conclude that, for 
the election systems $\electwo$ built in the proofs of 
many of the cases of Theorem~\ref{t:eight-part},
$\alwayswinners(\electwo)$ satisfies each %
theorem in the model where there is always a runoff,
and has the property that it always has a winner (when the 
candidate set is nonempty).
Since our standard model and the model where there is always a runoff are
the same for election systems that always have a winner,
$\alwayswinners(\electwo)$ also satisfies the conditions of the
theorem in our standard model.

However, we comment that the above transformation will be
useful in the exact same way for many types of manipulative attacks
other than the three discussed above.  It in fact will similarly work
(keeping in mind that we are always in the so-called nonunique-winner model---aka
the co-winner model---which focuses on whether a given candidate is/is not 
\emph{a} winner)
for all standard types of voter control
(adding/deleting/partitioning), all standard types of manipulation,
and all standard types of bribery.  Thus, the above transformation
goes quite far in paving over the divide between those who feel that
requiring nonempty winner sets is 
unnatural 
and those who feel that
failing to require nonempty winner sets is
unnatural.
}%

\alwayswinnersconstruction

\subsection{Specific Voting Rules and Scoring Protocols}

The following result provides---if we keep in mind 
that 
it is well-known that 
\cucm{Veto} is in $\p$---a natural, unweighted 
case where the classic manipulation problem
is simple but the runoff and revoting runoff versions are hard.
Even 
Narodytska and Walsh's~\cite{nar-wal:c:two-stage}
work, in which they allowed themselves the freedom to use 
different systems in the first and second round, did not obtain
any natural, unweighted example of runoffs or revoting runoffs 
increasing complexity.

\begin{theorem}\label{t:veto}
\cucmrunoff{Veto} and 
\cucmrevoting{Veto} are each $\npc$.
\end{theorem}

\begin{proofs}
We will reduce from the
well-known NP-complete Exact Cover by 3-Sets Problem (X3C):
Given a set $B = \{b_1, \ldots, b_{3k}\}$,
and a collection
${\cal S} = \{S_1, \ldots, S_n\}$ of three-element subsets of $B$, we
ask if ${\cal S}$ has an exact cover for $B$, i.e., if there exists a
subcollection
${\cal S'}$ of ${\cal S}$ such that every element of $B$ occurs in exactly
one member of ${\cal S'}$.
Without loss of generality, we assume that  $k \geq 1$ and
$n \geq 3$.  We will denote which elements of $B$ are in 
a given $S_i$ by 
some new ${i_j}$ variables:
$S_i = \{b_{i_1}, b_{i_2}, b_{i_3}\}$.

Since \cucm{Veto} is in P (simply greedily veto all 
candidates that score higher than $p$), the only place where hardness
can come in is in the selection of the set of winners in the 
initial round.  

Our election has the following set $C$ of candidates:
$p$ (the preferred candidate), 
$b_1, \ldots, b_{3k}$ and $s_1, \ldots, s_n$ 
(candidates corresponding to the X3C instance), 
$r_1, \ldots, r_k$ (candidates that will be vetoed in the runoff),
$d$ (a buffer candidate),
and $\ell$ (a candidate that always loses in the initial round).
We have $k$ manipulators.
We have the following nonmanipulators.
($\cdots$ in a vote denotes that the remaining candidates are in arbitrary
order.)
\begin{itemize}
\item
For every $i, 1 \leq i \leq n$, 
\begin{itemize}
\item
one nonmanipulator voting \hfill  $\cdots > p > b_{i_1} > s_i$,
\item
one nonmanipulator voting \hfill $\cdots > p > b_{i_2} > s_i$,
\item
and one nonmanipulator voting \hfill $\cdots > p > b_{i_3} > s_i$.
\end{itemize}
\item
Three nonmanipulators voting \hfill  $\cdots > p$.
\item
For every $c \in B \cup
\{r_1, \ldots, r_k\} \cup \{d\}$, 
three nonmanipulators voting \hfill $\cdots > p > c$.
\item
One nonmanipulator voting \hfill $\cdots > p > \ell$.
\item
For every $i, 1 \leq i \leq n$, 
one nonmanipulator voting \hfill $\cdots > p > d > s_i > \ell$.
\end{itemize}
Note that every candidate other than $\ell$ receives three vetoes from the
nonmanipulators in the initial round.  And
$\ell$ receives $n+1 > 3$ vetoes from the nonmanipulators.

Let ${\cal S'} = \{S_{j_1}, \ldots, S_{j_k}\} \subseteq {\cal S}$
be an exact cover for $B$.
 For $1 \leq i \leq k$, let the $i$th manipulator vote
$\cdots > r_i > s_{j_i}$.  We claim that $p$ is a winner of the
overall election (even without revoting).
It is immediate that the winner set of the initial round is
$C - \{\ell\} - \{s_{j} \ | \ S_j \in {\cal S'}\}$.
Since $\ell$ does not participate in the runoff,
$p$ gains one veto from the nonmanipulator voting
$\cdots > p > \ell$.  Since every candidate in 
$B \cup \{r_1, \ldots, r_k\} \cup \{d\}$
participates
in the runoff, this is the only veto that $p$ gains.
Since $\ell$ does not participate in the runoff,
each  $s_i$ that participates in the runoff gains one veto
from the nonmanipulator voting $\cdots > d > s_i > \ell$.
$d$ gains $k \geq 1$ vetoes from the
nonmanipulators voting $\cdots > d > s_i > \ell$ such that $S_i \in {\cal S'}$
and every $b \in B$ gains one veto from the
nonmanipulator voting $\cdots > p > b > s_i$ such that $b \in S_i$ and
$S_i \in {\cal S'}$.  Every candidate $r_i$ gains a veto from the
manipulator voting $\cdots > r_i > s_{j_i}$.  It follows that $p$
is a winner in the runoff.

For the converse, we will show the manipulations
described above are
the only way to make $p$ a winner.  Suppose the manipulators can
vote (in the initial round and the runoff) in
such a way that $p$ becomes a winner of the overall election.
Recall that in the initial round, 
every candidate other that $\ell$ receives three vetoes from the
nonmanipulators and that $\ell$ receives $n+1 > 3$ vetoes.
Since there are $k$
manipulators, $\ell$ does not participate in the runoff and
at most $k$ other candidates (the ones vetoed by a manipulator) do
not participate in the runoff.  Since $\ell$ does not participate in
the runoff, $p$ gains one veto from the nonmanipulator voting
$\cdots > p > \ell$.

Suppose there is a candidate $c \in
B \cup \{r_1, \ldots, r_k\} \cup \{d\}$ that does
not participate in the runoff.
Then $p$ gains three vetoes from the nonmanipulators
voting $\cdots > p > c$, and thus
$p$ receives at least seven vetoes in the runoff.
There are at least $2k$ candidates from $B$ that participate in
the runoff and each of these candidates is vetoed three times
in the initial round and does not gain any vetoes from deleting $\ell$.
Since $p$ receives at least seven vetoes in the runoff,
each candidate in $B$ that participates in the runoff
needs to gain at least four vetoes, so these candidates need
to gain a total of at least $8k$ vetoes.  But the most vetoes that
these candidates can gain in total is three vetoes for each candidate $s_i$ that
does not participate in the runoff
plus $k$ vetoes from the manipulators.  Since fewer than
$k$ $s_i$ candidates do not participate in the runoff,
the $B$ candidates that participate in
the runoff gain a total of at most $4k$ vetoes, which is not enough.

It follows that the only candidates other than $\ell$ that do not
participate in the runoff are $s_i$ candidates.  Note that candidates in 
$\{r_1, \ldots, r_k\}$ will not gain vetoes from the nonmanipulators in
the runoff, and
so each manipulator needs to veto exactly one $r_i$
in the runoff.  To make sure
that every candidate $b \in B$ gains at least one veto, we need to
delete a set of $s_i$ candidates corresponding to a cover.  Since we 
can delete at most $k$ such candidates, these candidates
will correspond to an exact cover.%
\end{proofs}

It is easy to argue, in contrast with the result of
Theorem~\ref{t:veto} regarding Plurality's close cousin Veto, 
that Plurality is
easy, even in the weighted case, since throwing all one's votes
to $p$ is always optimal.

It is important to make the distinction between our result
that \cwcmrunoff{Plurality} and
\cwcmrevoting{Plurality} are each in $\p$ and Narodytska and Walsh's
claim that computing a weighted coalition manipulation for Plurality
with Runoff is NP-hard with or without
revoting~\shortcite{nar-wal:c:two-stage}.
For their results
they use the definition \xtheny{TopTwo}{Plurality}, instead of the 
arguably more natural 
approach of using the same system at each stage, which is what we 
are considering here.

\begin{theorem}\label{t:plurality}
\cwcmrunoff{Plurality} and 
\cwcmrevoting{Plurality} are each in $\p$.
\end{theorem}

We mention the following result, which holds because 
by brute-force partitioning of the integer $\|V\|$ into
at most $(\|C\|!)^2$ named buckets (one for each pair of possible votes,
though a second-round decrease in candidates could make the numbers
even smaller than this), one can solve even the revoting runoff
manipulation question (and of course the same holds for plain runoffs).

\begin{theorem}\label{t:brute}
For any election system $\elec$ having a $\p$ winner problem,
and for any integer $k$, 
\cucmrevoting{$\elec$} restricted to $k$ candidates 
is in $\p$.
\end{theorem}

The following claim transfers to our two problems the dichotomy result
for scoring protocols known for the nonrunoff 
case.

\begin{theorem}\label{t:plur:triv}
For every scoring protocol $\elec$, 
\cwcmrunoff{$\elec$} and 
\cwcmrevoting{$\elec$} are in $\p$ if $\elec$ is 
Plurality or Triviality
 (or a direct transform of one of those, in 
a sense that can be made 
formal, see \cite{hem-hem:j:dichotomy}), and otherwise are $\npc$.
\end{theorem}

\begin{proofs}
For Plurality, this follows from Theorem~\ref{t:plurality} and
for Triviality, this is trivial.  For every other weighted scoring
protocol $\elec$,
Hemaspaandra and 
Hemaspaandra~\shortcite{hem-hem:j:dichotomy} give a reduction $f$ from
the NP-complete problem Partition to
\cwcm{$\elec$} with the property that for all
$x$, if $x \in $ Partition, then $p$ can be made the unique winner in $f(x)$,
and if $x \not \in $ Partition, then $p$ can not be made a winner in
$f(x)$.  So, if $x \in $ Partition, then $p$ can be made the unique winner
of the initial round, and thus the unique winner of the overall election.
And if $x \not \in $ Partition, then $p$ will never make it to the final
round.~\end{proofs}

The case of just one  manipulator is a natural and 
important case.
It also can often be surprisingly well handled,
thanks to the lovely result---initially due for the 
unweighted case to the seminal 
work of 
Bartholdi, Tovey, and Trick~\shortcite{bar-tov-tri:j:manipulating}
and 
since then much extended---that the natural (p-time) greedy manipulation 
algorithm (giving one's highest point value to $p$ and then
giving, in turn, the highest remaining value to 
the candidate who has the lowest point total among those not yet 
assigned points by the manipulative voter)
is optimal (i.e., finds a successful 
manipulation when one exists) for
both weighted and unweighted 
scoring protocols, for the case when there is just one manipulator.
The following 
theorem implies %
that that result does not carry 
over to runoff elections.

\begin{theorem}\label{t:one-evil}
The standard one-manipulator p-time greedy algorithm for scoring
protocols is not optimal for 
\cucmrunoff{HalfApproval} and 
\cucmrevoting{HalfApproval},
restricted
to at most one manipulator.
\end{theorem}

\def\oneevilproof{\begin{proofs}
Consider the election with candidate set $\{p,a,b,c\}$, a nonmanipulator
voting $a > p > b > c$, a nonmanipulator voting
$a > b > p > c$, and one manipulator.
The scores of $p, a, b, c$ from the nonmanipulators
are $1, 2, 1, 0$.  The greedy algorithm would give the following
vote for the manipulator: $p > c > b > a$. 
Then $p$ and $a$ are the winners of the initial round, and there
is no way for $p$ to win the runoff.  However, if the manipulator
votes $p > b > c > a$, then $p$, $a$, and $b$ are the winners of
the initial round, and $p$ is a winner of the runoff (even without
revoting).~\end{proofs}}

\oneevilproof

Theorem~\ref{t:veto} gave a case where a simple-to-manipulate
unweighted scoring protocol became hard for runoffs, with or without
revoting.  The following result gives a new example of runoffs
increasing complexity, this time for the one-manipulator case.  
It is
natural to wonder whether the following theorem itself implies
Theorem~\ref{t:one-evil}.
The answer is 
that the following theorem does not imply that, but it does imply something
a bit weaker,
namely, it says that
Theorem~\ref{t:one-evil}
holds unless $\p = \np$.  (Of course, 
Theorem~\ref{t:one-evil} holds absolutely;  it doesn't require 
a $\p \neq \np$ hypothesis.) %
\cucm{HalfApproval} for one manipulator is clearly in $\p$---for 
example by the greedy algorithm we mentioned 
above---and so the following result does
express a raising of complexity.
Even \cucm{HalfApproval} (i.e., with an unbounded number of
manipulators) is in $\p$ (this follows from the argument
of~\cite[Corollary 4.2]{con-pro-xia:c:scheduling-manipulation}).
So this gives even more of a contrast between classic and 
runoff manipulation complexity.

\begin{theorem}\label{t:half}
\cucmrunoff{HalfApproval} and 
\cucmrevoting{HalfApproval} are each $\npc$, even when restricted
to having at most one manipulator.
\end{theorem}

\def\halfproof{\begin{proofs}
This construction operates similarly to the construction from
Theorem~\ref{t:veto}.
Note that we have fewer candidates in the runoff than in the initial
round, and so in contrast to the proof of Theorem~\ref{t:veto},
the manipulator has fewer vetoes to contribute in the runoff and some
candidates may have fewer vetoes in the runoff than they had in the initial
round. We must be careful of how these two points affect our construction.
We reduce from Exact Cover
by 3-Sets (X3C): Given a set $B = \{b_1, \ldots, b_{3k}\}$, and a
collection ${\cal S} = \{S_1, \ldots, S_n\}$ of three-element
subsets of $B$, we ask if ${\cal S}$ has an exact cover for $B$, i.e.,
if there exists a subcollection ${\cal S'}$ of ${\cal S}$ such that every
element of $B$ occurs in exactly one member of ${\cal S'}$. We will denote
which elements of $B$ are in a given $S_i$ by some new $i_j$ variables:
$S_i = \{b_{i_1}, b_{i_2}, b_{i_3}\}$. Without loss of generality we
assume that $n \geq 3$ and $n \geq k \geq 1$.

We pad the %
election from the proof of Theorem~\ref{t:veto}. %
We have the following candidates: $p$ (the preferred candidate),
$B = \{b_1, \dots, b_{3k}\}$ and $S = \{s_1, \dots, s_n\}$
(corresponding to the X3C instance), 
$R = \{r_1, \ldots, r_{n + 2k}\}$ (candidates that will pad the votes in
the runoff), $d$ (a buffer candidate), and
$L = \{\ell_1, \ldots, \ell_{2n + 5k}\}$ and $\hat{\ell}$ (candidates that
will lose in the initial round).
Summarizing, this gives us the candidate set
$C = \{p,d,\hat{\ell}\} \cup B \cup S \cup R \cup L$.
We have $\|C\| = 4n + 10k + 3$, so each vote approves
$\left \lceil \|C\|/2 \right \rceil = 2n + 5k + 2$ candidates
and vetoes $2n + 5k + 1$ candidates. Observe that the number of vetoes
contributed by a vote is equal to $\|L\| + 1$, which is crucial to pad our construction.
We have the following nonmanipulators.
(A set in the specification of a vote, e.g., $R$ in the first vote specified
below, denotes that candidates in that set are in arbitrary order in that
part of the vote.)
\begin{itemize}
  \item For every $i$, $1 \leq i \leq n$,
    \begin{itemize}
      \item one nonmanipulator voting \hfill
            $\cdots > B - \{b_{i_1}\} > p > b_{i_1} > R > s_i > L$,
      \item one nonmanipulator voting \hfill
            $\cdots > B - \{b_{i_2}\} > p > b_{i_2} > R > s_i > L$,
      \item one nonmanipulator voting  \hfill
		    $\cdots > B - \{b_{i_3}\} > p > b_{i_3} > R > s_i > L$,
	  
	  \item and one nonmanipulator voting  \hfill
		   $\cdots > B > d > R > s_i > \hat{\ell} > L$.
	\end{itemize}
  \item For every $i$, $1 \leq i \leq 3k$,
    \begin{itemize}
      \item two nonmanipulators voting \hfill $\cdots > B - \{b_i\} > p > R > b_i > L$.
    \end{itemize}
  \item For every $i$, $1 \leq i \leq n + 2k$,
    \begin{itemize}
      \item two nonmanipulators voting \hfill
                    $\cdots > B > p > d > R - \{r_i\} > r_i > L$.
    \end{itemize}
  \item Two nonmanipulators voting \hfill $\cdots > B > R > d > L$.
  \item Three nonmanipulators voting \hfill $\cdots > B > R > p > L$.
  \item One nonmanipulator voting \hfill $\cdots > B > R > p > \hat{\ell} > L$.
\end{itemize}

The votes of the nonmanipulators result in the following vetoes for
each of our candidates. (We count vetoes since the main idea of this proof
is similar to the proof of Theorem~\ref{t:veto} and since we specify enough of
each of the votes to clearly show all possibilities for the initial round
and the runoff.) Recall that the number of vetoes contributed by %
each vote in the initial
round is $\|L\|+1$.
\begin{itemize}
  \item $p$ has three vetoes.
  \item Each candidate in $S$ has three vetoes.
  \item Each candidate in $B \cup R \cup \{d\}$ has two vetoes.
  \item Each candidate in $L$  has more than three vetoes.
  \item $\hat{\ell}$ has more than three vetoes (since $n \geq 3$).
\end{itemize}

Let ${\cal S'} \subseteq {\cal S}$ be an exact cover for $B$.
We will show that if the manipulator votes
\[\cdots > L > \{s_i~|~S_i \in {\cal S'}\} > R > d > B\]
then $p$ will be a
winner of the overall election.
Note that the candidate set of the
runoff is %
$C' = \{p,d\} \cup B \cup \{s_i~|~S_i \not \in {\cal S'}\} \cup R$
and $\|C'\| = 2n+4k+2$.
Again we have a set of candidates that is crucial to pad the votes in the
runoff. Now $\|R\|+1$ is the number of vetoes in the runoff not unlike when
$\|L\|+1$ was the number of vetoes in the initial round.
In the runoff, because of the absence of the candidates in
$L \cup \{\hat{\ell}\}$
and the candidates corresponding to the exact cover,
and because of the vote of the manipulator,
the remaining candidates have the following vetoes.
\begin{itemize}
  \item $p$ has four vetoes. %
  \item Each candidate in $S$ that participates in the runoff has
        four vetoes. %
  \item Each candidate in $B$ has four vetoes. %
  \item Each candidate in $\{d\} \cup R$ has more than four vetoes.
\end{itemize}
So, $p$ is a winner in the runoff (even without revoting). 
For the converse suppose
that the manipulator votes (in the initial round and the runoff) such that
$p$ is a winner of the overall election.
We will show that this is only possible if
${\cal S}$ has an exact cover.

Clearly, each candidate in $L \cup \{\hat{\ell}\}$ will not participate in the
runoff since the manipulator can only possibly add one additional veto
to each candidate. So there is no way for any of the candidates
in $B \cup R \cup \{d\}$ to have more than three vetoes in the initial round.
Let $C'$ be the set of candidates that participate in the runoff.
Note that $C' = \{p,d\} \cup B \cup \widehat{S} \cup R$, where
$\widehat{S} \subseteq S$.  

If $||C'|| < 2n + 4k + 2$, then
each voter vetoes at least one and at
most $n + 2k = ||R||$ candidates in the runoff.
This causes $d$ to receive two vetoes while
$p$ receives four vetoes from the nonmanipulators in the runoff.
Since there is only one manipulator, $p$ will not win the runoff.

If $||C'|| \geq 2n+4k+4$, then
each voter vetoes at least $||R|| + 2$ and
certainly no more than $||R|| + ||B||$
candidates in the runoff.
Note that $\widehat{S} \neq \emptyset$ and that
every candidate in $\widehat{S}$ is vetoed four times by the
nonmanipulators in the runoff while $p$ clearly has
more than five vetoes.
Since there is only one manipulator, $p$ will not win the runoff.

It remains to handle the case where
$2n + 4k + 2 \leq ||C'|| \leq 2n+4k+3$.  In this case,
each voter vetoes $||R|| + 1$ candidates in the runoff.
$p$ has four vetoes from the nonmanipulators in the runoff.
So, every $b \in B$ needs at least three vetoes from the 
nonmanipulators in the runoff.  The only way $b$ can get more than two vetoes
from the nonmanipulators is if for some $i$,
$b \in S_i$ and $s_i$ does not participate in the runoff.
It follows that $S - \widehat{S}$ 
corresponds to a cover for $B$.
Since $||S - \widehat{S}|| \leq k$, it follows that
$S - \widehat{S}$  corresponds to an exact cover for~$B$.~\end{proofs}

}

\halfproof

Theorems~\ref{t:veto} and~\ref{t:half} give examples of natural
unweighted systems where runoffs increase the complexity
of manipulation.  What about the weighted case?  This is harder, since
there are far fewer examples of natural weighted election systems for
which manipulation is easy.  And those examples tend to be so easy
that they remain easy with runoffs.  For example, 
weighted manipulation for scoring protocols is easy if and only if the system
is Plurality or Triviality, and those remain easy
(see Theorem~\ref{t:plurality}).
Narodytska and Walsh~\shortcite[Proposition 7]{nar-wal:c:two-stage} show that
\xtheny{Condorcet}{Plurality} with weighted votes is NP-complete
with three or more candidates.  Their definition of Condorcet selects 
as winners the Condorcet winner (i.e., the candidate that beats every
other candidate in a pairwise election) and all candidates if there
is no Condorcet winner.  However, it is immediate from their proof
that \cwcm{Condorcet} (in the definition of~\cite{nar-wal:c:two-stage})
with at least three candidates is already NP-complete,
and so the second round doesn't increase the complexity.

Narodytska and Walsh~\shortcite{nar-wal:c:two-stage} also 
show that \xtheny{TopTwo}{Plurality} with weighted votes is NP-complete,
even when restricted to three candidates.  
This results holds with or without revoting.
TopTwo selects the two candidates with the highest plurality scores
(ties are broken lexicographically, and the candidates are renamed
so that the preferred candidate is lexicographically first).
It is easy to see
that \cwcmrunoff{TopTwo} and \cwcmrevoting{TopTwo} are easy, since 
voting for $p$ is always optimal.  
Now consider the election system TopTwo$_{\mathrm{m}}$, where the winner
is the Majority winner (if it exists) and otherwise the two candidates
with the highest Plurality scores
(ties are broken lexicographically, with the preferred
candidate viewed as being lexicographically first).
\cwcm{TopTwo$_{\mathrm{m}}$} is clearly in $\p$; the optimal strategy
for the manipulators is to vote for $p$. 
It is easy to see that \cwcmrunoff{TopTwo$_{\mathrm{m}}$} is the
same problem as weighted manipulation of \xtheny{TopTwo}{Plurality} 
(sometimes called Plurality with Runoff in~\shortcite{nar-wal:c:two-stage})
and that \cwcmrunoff{TopTwo$_{\mathrm{m}}$} is the
same problem as weighted manipulation of \xtheny{TopTwo}{Plurality} 
with runoff.  Thus the following holds.

\begin{theorem}\label{t:toptwo}
    \cwcm{TopTwo$_{\mathrm{m}}$} is in $\p$, while
    \cwcmrunoff{TopTwo$_{\mathrm{m}}$} and
    \cwcmrevoting{TopTwo$_{\mathrm{m}}$} are each $\np$-complete, even when
    restricted to three candidates.
\end{theorem}
Is there an example of an election system \emph{that is already in the
literature} for which runoffs increase the complexity of 
weighted manipulation?
There is one known case where weighted manipulation is easy, but
in a very nontrivial way.  This is \cwcm{Llull}, restricted to
four candidates~\cite{fal-hem-sch:c:llull4}.  This is in contrast to 
Copeland$^\alpha$ for other values of $\alpha$ (recall that 
Llull is Copeland$^1$), which are hard
for three or more
candidates~\cite{fal-hem-sch:c:copeland-ties-matter,fal-hem-sch:tOUTbyCONF:copeland-ties-matter}.
The reason that we don't get hardness
in the same way in the case for Llull, is that we get hardness by
enforcing ties (which allows us to encode Partition), but in the 
case of Llull, a tie is never better than a non-tie for making $p$
a winner.  However, if we have a second round, it does not just matter
if $p$ wins, but also if other candidates win, since these candidates
participate in the second round.  It turns out that to make candidates
other than $p$ win, we sometimes need to enforce a tie, which allows
us to encode Partition, and we get NP-completeness.  This already
happens for four candidates,
and so we have,
in four-candidate weighted Llull, found an example of a case where
runoffs make the problem go from easy to hard.\footnote{Runoffs do not
increase the complexity for three-candidate weighted Llull, since
in that case $p$ can be made a winner
of the initial round if and only if $p$ can be made a winner of the
election with runoff.
If this were not true, there would be 
a candidate $a$ such that $a$ defeats $p$ in their pairwise election
and the winners of the initial round are $p$ and $a$.
But then the score of $p$ in the initial round is $1$, and the
only way $p$ can be a winner of the initial round is if the
third candidate, $b$, defeats $a$ in their pairwise election.  But
then $b$ is also a winner of the initial round.}

\begin{theorem}\label{t:llull}
\cwcmrunoff{Llull} and 
\cwcmrevoting{Llull} are each $\npc$, even when restricted to 
four candidates.
\end{theorem}

\def\llullproof{\begin{proofs}
When looking at Llull elections, it is often convenient to think of an election
as its induced weighted majority graph.  Given an election $E = (C,V)$,
$E$'s induced weighted majority graph is the directed graph that has
$C$ as its vertices and for each pair of vertices $c, d$, there
is an edge from $c$ to $d$ with weight $w > 0$ if 
$c$ beats $d$ by a margin of $w$ in the pairwise election between $c$ and $d$.
If we leave off the weights, we get the induced majority graph.
Note that we can determine the winners of a Llull election from
its induced majority graph.

As in Section~\ref{sss:weighted}, we
will reduce from the well-known NP-complete problem Partition: Given a
nonempty set of positive integers $k_1, \ldots, k_t$ that sums to $2K$,
we ask if there exists a subset that sums to $K$.
Let $C = \{p,a,b,c\}$.  Construct a set of nonmanipulators $V_1$ such that 
the induced weighted majority graph of $(C,V_1)$ looks like this:

\begin{figure}[H]
    \centering
    \includegraphics{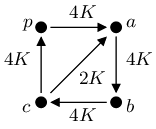} 
\end{figure}

Such a $V_1$ exists (since all weights have the same parity)
and can be computed in polynomial time using
McGarvey's construction~\cite{mcg:j:election-graph}.
The manipulators have weights $k_1, \ldots, k_t$.
Since the total weight of the manipulators is $2K$, 
no matter how the manipulators vote, the induced majority graph contains
the cycle $p \rightarrow a \rightarrow b \rightarrow c \rightarrow p$
and does not contain the edge $a \rightarrow c$.

If $k_1, \ldots, k_t$ can be partitioned into two sets of equal weight
(i.e., weight $K$), the manipulators vote so that $K$ of the manipulator
weight votes $p > b > a > c$ and
$K$ of the manipulator weight votes $b > p > a > c$.
This gives the following induced majority graph (drawing an undirected
edge for a tie):

\begin{figure}[H]
    \centering
    \includegraphics{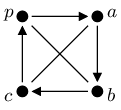}
\end{figure}

Note that all candidates have a score of 2.  So, all candidates proceed
to the second round. Without revoting, all candidates 
win the second round.  So, even without revoting $p$ is a winner of
the election.

For the converse, suppose that $p$ is a winner of the election.
$c$ always makes it to the second round.
If the set of candidates in the second round is $\{p,c\}$ or
$\{p,a,c\}$, $c$ is the unique winner.
So, in order for $p$ to be a winner, both $p$ and $b$ need to make
it to the second round.  This implies that $p$ and $b$ tie in 
the first round.
Consider the set of weights of the manipulators for which $p > b$ in the
first round.  This set sums to $K$ and so
we have partition.~\end{proofs}

Note that the construction above also gives a natural election system
for which it is not always better for the manipulators to put $p$
first in the first round (although of course doing so would be optimal
in classic manipulation).
} %

\llullproof %

We realize that 
upon seeing 
results such as 
Theorems~\ref{t:veto}--\ref{t:plurality}, 
\ref{t:plur:triv}, and 
\ref{t:half}--\ref{t:llull} 
it might be natural to wonder whether NP-hardness of 
\cwcmrunoff{$\elec$} automatically implies NP-hardness 
of \cwcmrevoting{$\elec$}; Theorem~\ref{t:eight-part}
however shows
that no such universal implication holds.
\section{Conclusions and Open Problems}
This paper has explored the relative manipulation complexity of runoff
elections, with and without revoting.  We have seen that there is no
general relation between the manipulation complexity of either of
those with each other or with the manipulation complexity of the
underlying election system.  Sometimes revoting can even lower
complexity, for example.  Yet for the natural, concrete systems we
studied, runoffs and revoting runoffs never lowered complexity and
sometimes raised complexity.

Important open directions include the study of runoffs and 
revoting runoffs for the case of bribery 
rather than manipulation, for which we have 
some preliminary results;
seeking to find a natural 
system for which the complexity 
of \cwcmrunoff{$X$} and \cwcmrevoting{$X$} differ (even if 
one is allowed to use different first- and second-round systems,
this question is open in the literature);
and the study of what role heuristics, especially in light of 
Theorems~\ref{t:one-evil}
and~\ref{t:half}, can play.

\bibliographystyle{alpha}
\def\twoofeightproofsketches{%
\begin{theorem}\label{t:npc:p:npc}There exists an election system $\elec$, whose winner problem
is in $\p$ such that \cucm{$\elec$} is $\npc$, \cucmrunoff{$\elec$} is in $\p$
and \cucmrevoting{$\elec$} is $\npc$.\end{theorem}
\begin{proofsketch}
Recall that we promised that all our election systems would always
have at least one winner (if there is at least one candidate).  The
election system that we build in this theorem will satisfy that
also.  However, along the path to building such a system $\elec$, we will
first build an election system $\electwo$ that is allowed to have everyone
lose, and that satisfies the theorem.  
And then we'll show how to transform $\electwo$ into a system $\elec$ that
ensures that there is always at least one winner, and yet is such that 
$\elec$ 
still establishes our theorem.

Let $f'(\ell)$ denote the number of dummy candidates needed so that an ordering
over $f'(\ell)$ candidates can express clearly at least $\ell$ possibilities. So, $f'(\ell)$
can be taken to be the least integer $k$ such that $k! \ge \ell$.
Let $\electwo$ be defined as follows:

If $\|V\| \neq 1$ or $\|C\| = 1$ or $\|C\| = 3$ then everyone loses.
Otherwise, we'll handle things based on the number of candidates and the single
voter $v$, as described below.

If $\|C\| = 2$ and the lexicographically larger candidate is more preferred
than the lexicographically smaller candidate by $v$, then everyone wins,
else everyone loses.

If $\|C\| \geq 4$ and the lexicographically smallest candidate's name
encodes a formula $\psi$, with $\numvars(\psi) = k$, and $\|C\| \ge f'(2^k)+1$
(the ``$+1$'' is as we'll not involve the lexicographically smallest
candidate in encoding the assignment)
and $v$'s vote on all candidates
other than the lexicographically smallest candidate encodes an assignment
that satisfies $\psi$ and
the lexicographically smallest candidate is more preferred than the
lexicographically largest candidate by $v$, then
the lexicographically largest and smallest candidates win, else
everyone loses.

\noindent
That completes the specification of $\electwo$.

Now we explain why this system 
$\electwo$ meets all the requirements of the theorem.
\begin{itemize}
    \item Clearly the winner problem for $\electwo$ is in $\p$.
    \item To show that \cucm{$\electwo$} is $\npc$ observe that $\psi \in 
          \sat$ reduces to \cucm{$\electwo$} for the candidate set: $p$
          encoding $\psi$ and
          $\max(3, f'(2^k))$ dummy candidates (where $\numvars(\psi) = k$)
          all lexicographically
          larger than $p$. Let the preferred candidate be $p$ and let there
          be 0 nonmanipulative votes and 1 manipulative voter.
    \item To show that
          \cucmrunoff{$\electwo$} is in $\p$ observe that 
          in the first round we always have 0 or 2 winners.
          However, when we have 2 winners (and more than 2 candidates),
they always lose in the second
          round unless $v$ can change her vote.
So in the second round scheme
          (runoff without revoting), no one ever can win if we have
more than two candidates.
    \item To show that \cucmrevoting{$\electwo$} is $\npc$, we can use the same
          example that we used to show that \cucm{$\electwo$} is $\npc$. Here, it
          works by, when the $\psi$ encoded by the lexicographically smallest candidate 
          is satisfiable, $v$ in the first round conveys
          (in the dummy candidates)
          a satisfying assignment, and it puts the lexicographically smallest 
          candidate as its most preferred candidate.
          In the second round it puts the lexicographically smallest
          candidate as its least preferred candidate.
          (So we'll go to the $\|C\| = 2$ case 
          of $\electwo$ in the second round and we will have 2 second round 
          winners, including the lexicographically smallest candidate.)
\end{itemize}
So $Y$ satisfies the theorem.  But $Y$ for some inputs 
may have no winners, and we promised to ensure in our constructions that
there will always be at least one winner (when there is at least one 
candidate).  We handle this as follows.  Let $X = \alwayswinners(Y)$,
where $\alwayswinners$ is as described immediately after this proof.
Then $X$, which always has at least one 
winner (when there is at least one candidate),
also satisfies the conditions of the theorem, and so we are done.

This completes the proof sketch.~\end{proofsketch}

Within the proof, we mentioned a transformation $\alwayswinners$.  
We will give it here.  The reason we give it outside of the above
proof sketch is that this transformation is of more general interest
and applicability.
Indeed, we'll use it again almost immediately, 
in the proof of 
Theorem~\ref{t:npc:npc:p}.  What this transformation will do is 
it will take an election system and will transform it into a new 
election system that will (except when the candidate 
set is empty) always have at least 
one winner, yet that is so closely related to the original election system
that the complexity of the original system, 
with respect to the three problems we are concerned with here (and in fact,
with respect to many, many other types of manipulative actions) 
is unchanged.  

This transformation thus relatively broadly addresses a persistently 
annoying issue.  Many people, including the authors, feel that it is 
ugly and asymmetric to require nonempty winner sets yet to allow 
all of $C$ to be the winner set.  And indeed a number of papers in 
\emph{computational} social choice 
do allow the winner set to be 
any member of $2^C$.  On the other hand, traditionally in 
\emph{social choice}, the requirement that the winner set be nonempty
is part of the definition of elections.  Our transformation 
shows that the difference in models isn't as large as one might 
think; one can often adopt the more symmetric model, yet by this 
transformation one will know that one's results also hold in the 
more restrictive model.  

We now give the transformation $\alwayswinners$.  First let us describe
informally how it works, and then we'll describe it and its properties
more formally.  Let $\elece$ be the election system we want to
transform under $\alwayswinners$.  Informally, suppose that we have a
candidate name, $new$, that is not part of our universe of legal
candidate names.  (Of course this is untrue; the universe of names is
as it is.  But please indulge us for a few more lines, and we'll then 
avoid this problem 
in our more formal approach.)  
Then under $\alwayswinners$, if $new$ is not an input
candidate then all candidates win, and if $new$ is an input candidate
then $new$ wins and also all candidates win who under $\elece$ would
win if in our input election $new$ is removed and all votes are masked
down to remove $new$.  Note that this system always has a winner (if 
the set of candidates is not empty).  Also, it can be easily 
seen to retain
most of the manipulative-action 
complexities of $\elece$, as we'll discuss later.

As admitted above, we can't just make a new name appear.  But we can get 
the same effect formally, by shifting all the names in the universe up 
by one spot to open up a space for our new name, and then when we're 
using those other names in simulating the original system, by shifting 
them back
down again.  And that is precisely what we will do.  

So we now more carefully and correctly specify the transformation
$\alwayswinners$.
Let $\elece$ be 
an election system (that perhaps has no winners even on inputs
on which $C$ is nonempty).  Let the set of legal names for 
candidates (i.e., the set of all strings) be enumerated in 
lexicographic order by $s_0, s_1, s_2, \ldots$ (e.g.,
``$\epsilon$, 0, 1, 00, 01,~$\ldots$'' if names are taken to be 
binary strings).  Let $\texttt{++}$
denote a one-step increase in this order, i.e., $s_i{\texttt{++}} = s_{i+1}$.
The $\texttt{++}$ operator naturally applies to sets of candidates, namely
as defined by 
$A{\texttt{++}} = \{ a\texttt{++} \condition a \in A\}$.  And for any set 
$A$ of candidates such that $s_0 \not\in A$, we similarly 
define the decrement of the set, namely by
$A\mbox{-\,-} = \{ a \condition a{\texttt{++}} \in A\}$.
On candidate set $C$ and voter set $V$, $\alwayswinners(\elece)$ does 
the following.  If $s_0 \not\in C$, then the winner set is $C$.
If $s_0 \in C$ then the winner set is $
\elece( 
(C - \{ s_0  \})\mbox{-\,-}, V')\texttt{++} \cup 
\{s_0\}$, 
where $V'$ is $V$ with $s_0$ masked 
out of each preference order and then each candidate name decremented
in each order and where $\elece(\hat{C},\hat{V})$ denotes the winner 
set, under $\elece$, of the election over candidate set $\hat{C}$
and voter set $\hat{V}$.

The crucial things to notice about $\alwayswinners(\elece)$ are 
the following, which hold for all election systems $\elece$ (including
ones that allow there to be no winners on some inputs for which
the input candidate set of the 
$\elece$ instance is nonempty).  
$\alwayswinners(\elece)$ always has at least one 
winner 
(when the input candidate set of the 
$\alwayswinners(\elece)$ instance is nonempty).  
For \cucm{$\elece$} (respectively, \cucmrunoff{$\elece$},
\cucmrevoting{$\elece$}) it holds that if the problem is in $\p$ 
then \cucm{$\alwayswinners(\elece)$} (respectively,
\cucmrunoff{$\alwayswinners(\elece)$},
\cucmrevoting{$\alwayswinners(\elece)$}) is in $\p$.
For \cucm{$\elece$} (respectively, \cucmrunoff{$\elece$},
\cucmrevoting{$\elece$}) it holds that if the problem is $\npc$ 
then \cucm{$\alwayswinners(\elece)$} (respectively,
\cucmrunoff{$\alwayswinners(\elece)$},
\cucmrevoting{$\alwayswinners(\elece)$}) is $\npc$.
Part of the easy task 
of seeing that these complexity connections hold is noticing 
that given an instance of one of these problems under $\elece$,
one can increment all candidate names both within the 
candidate set and the voter preferences, can then add in 
a new candidate $s_0$ and extend voter preferences arbitrarily 
to include that new candidate (e.g., putting it last in each 
voter's preferences),  and then we can note that a candidate $p$ 
can be made a winner 
in the initial election under $\elece$ exactly if 
$p\texttt{++}$ can be made a winner 
under $\alwayswinners(\elece)$ in the transformed election.  (And 
we mention in passing that the 
analogous claim holds for the so-called ``destructive'' case in 
which we seek to preclude $p$ from being a winner, though destructive 
cases are not a focus of this paper.)

The observations above are what we need to conclude that, for 
the $Y$ built in the above proof, $\alwayswinners(Y)$ satisfies the 
theorem, and has the property that it always has a winner (when the 
candidate set is nonempty).  

However, we comment that the above transformation will be
useful in the exact same way for many types of manipulative attacks
other than the three discussed above.  It in fact will similarly work
(keeping in mind that we are always in the so-called nonunique-winner model---aka
the co-winner model---which focuses on whether a given candidate is/is not 
\emph{a} winner)
for all standard types of voter control
(adding/deleting/partitioning), all standard types of manipulation,
and all standard types of bribery.  Thus, the above transformation
goes quite far in paving over the divide between those who feel that
requiring nonempty winner sets is 
unnatural 
and those who feel that
failing to require nonempty winner sets is
unnatural.

\begin{theorem}\label{t:npc:npc:p} There exists an election system $\elec$, whose winner problem
is in $\p$ such that \cucm{$\elec$} is $\npc$, \cucmrunoff{$\elec$} is 
$\npc$ 
and \cucmrevoting{$\elec$} is $\p$.\end{theorem}
\begin{proofsketch}
As in the proof of Theorem~\ref{t:npc:p:npc}, we will
first build an election system $\electwo$ that is allowed to have everyone
lose, and that satisfies the theorem and then
we let $X = \alwayswinners(Y)$,
where $\alwayswinners$ is as described immediately before this proof.

Let $f'(\ell)$ be as defined and used in the proof of Theorem~\ref{t:npc:p:npc}. 
The allowed (all others will cause everyone to lose) candidate types for $\electwo$ are described in
Table~\ref{allow:c:type}.

\begin{table}[t]
\centering
\def\arraystretch{1.5}
\begin{tabular}{  l | p{'220pt} } 
    \text{Candidate Form} & \text{Role in our proof} \\ \hline
    $\pair{1,1}$ & Seeks to make the first round hard.  \\
    $\pair{1,2}$ & Seeks to make the first round easy. \\ 
    $\pair{2,\psi}$ & Candidate coding a formula intended as
                 part of a hard first round. \\
    $\pair{3,\psi}$ & Candidate coding a formula intended as part
                of a hard second round. \\ 
    $\pair{4,\mbox{(any string)}}$ & ``Type-4'' dummy candidate, 
                    used to make votes so big as to encode assignments. \\ 
    $\pair{5,1},\pair{5,2}$ & Special dummy candidates to allow vote changes 
                to show through in some cases. \\ 
\end{tabular}
    \caption{{}Allowed candidate types for the election $\electwo$ used in the proof 
        of Theorem~\ref{t:npc:npc:p}\label{allow:c:type}.}
\end{table}

For our election $\electwo$, the candidate set is expected to be in one of the following
forms:
\begin{itemize}
    \item[(a)] $\pair{1,1},\pair{2,\psi}$, and enough
        type-4 dummy candidates so that a vote including
        them can encode an assignment to $\psi$, i.e., at
        least $f'(2^{{\numvars(\psi)}})$ dummy candidates.
    \item[(b)] $\pair{1,2},\pair{3,\psi},\pair{5,1},\pair{5,2}$, and enough
        type-4 dummy candidates so that a vote including
        them can encode an assignment to $\psi$, i.e., at
        least $f'(2^{{\numvars(\psi)}})$ dummy candidates.
    \item[(c)] $\pair{3,\psi},\pair{5,1},\pair{5,2}$, and enough
        type-4 dummy candidates so that a vote including
        them can encode an assignment to $\psi$, i.e., at
        least $f'(2^{{\numvars(\psi)}})$ dummy candidates.
\end{itemize}
Let $\electwo$ be defined as follows:

If $\|V\| \neq 1$ or $C$ is not of an expected form
then everyone loses.
Otherwise, we'll handle things as 
described below (note that in this case there is a single voter $v$).

If $C$ is of form (a) and the type-4 dummy candidates'
restriction of $v$'s vote encodes a satisfying assignment
to $\psi$, then $\pair{2,\psi}$ wins. Otherwise, everyone loses.

If $C$ is of form (b) and $\pair{5,2} > \pair{5,1}$ in $v$'s vote
then $\pair{3,\psi},\pair{5,1},\pair{5,2},$ and all of the type-4 dummy
candidates win.
In all other cases where $C$ is of form (b), everyone loses.
        
If $C$ is of form (c) and $\pair{5,1} > \pair{5,2}$ in $v$'s vote
or the type-4 dummy candidates' restriction of $v$'s vote encodes
a satisfying assignment to $\psi$, then $\pair{3,\psi}$ wins.
Otherwise, everyone loses.

Now we must explain why this system 
$\electwo$ meets all the requirements of the theorem.

\begin{itemize}
    \item Clearly the winner problem for $\electwo$ is in $\p$.
    \item \cucm{$\electwo$} is $\npc$ since to test if $\psi
         \in \sat$, we can ask if $\pair{2,\psi}$ can win the \cucm{$\electwo$} instance
          with
          candidate set \{$\pair{1,1},\pair{2,\psi},f'(2^{{\numvars(\psi)}})$
          type-4 dummy candidates\} and
          0 nonmanipulative votes and 1 manipulative voter.
    \item \cucmrunoff{$\electwo$} is $\npc$ since to test
          if $\psi \in \sat$, we can ask if $\pair{3,\psi}$ can win the
          \cucmrunoff{$\electwo$} instance
          with candidate set \{$\pair{1,2},\pair{3,\psi},\pair{5,1},\pair{5,2},
          f'(2^{\numvars(\psi)})$ type-4
          dummy candidates\} and 0 nonmanipulative votes
          and 1 manipulative voter. 

Note that we need $\pair{5,2} > \pair{5,1}$ in $v$'s vote in order for
          $\pair{3,\psi}$ to make it to the second round.
          If $\psi \notin \sat$, then $\pair{3,\psi}$ cannot win. Otherwise,
          if $\psi \in \sat$, voting so that $\pair{5,2} > \pair{5,1}$ in $v$'s
          vote and with the type-4 dummy order giving a satisfying
          assignment to $\psi$, makes
          $\pair{3,\psi}$ a winner.
    \item \cucmrevoting{$\electwo$} is in $\p$ by looking at each of the allowed
        candidate set forms:\hfill
        \begin{itemize}
            \item For form (a), everyone loses by the end of the second round.
            \item For form (b), when voter $v$ casts a vote where
                $\pair{5,2} > \pair{5,1}$
                in the first round and a vote where $\pair{5,1} > \pair{5,2}$
                in the second round then everyone wins who can possibly win.
            \item For form (c), everyone loses by the end of the second round.
        \end{itemize}
        All other cases have everyone lose immediately.
        Thus this problem is in $\p$.
\end{itemize}
This completes the proof sketch.~\end{proofsketch}
}

\end{document}